\documentclass[twocolumn, twocolappendix,times,tighten,10pt]{aastex631}
\usepackage{multirow}
\usepackage{amsmath}
\usepackage{booktabs}
\usepackage{floatrow}
\usepackage{graphicx}
\usepackage{verbatim}
\usepackage{amssymb}
\usepackage{bm}
\usepackage{hyperref}
\usepackage{xcolor}
\usepackage{comment}
\begin{document}


\title{Radiative Feedback in Population III Protostellar Growth: H\,\textsc{i} Shielding \& H\,\textsc{ii} Region Trapping}

\correspondingauthor{Avi Chen}

\author[0000-0002-8859-7790]{Avi Chen}
\affiliation{Department of Physics and Astronomy, Rutgers, The State University of New Jersey, 136 Frelinghuysen Rd, Piscataway, NJ 08854, USA \\}
\email{avi.chen@rutgers.edu}

\author[0000-0002-0311-2206]{Shyam H. Menon}
\affiliation{Center for Computational Astrophysics, Flatiron Institute, 162 Fifth Avenue, New York, NY 10010, USA \\}
\affiliation{Department of Physics and Astronomy, Rutgers, The State University of New Jersey, 136 Frelinghuysen Rd, Piscataway, NJ 08854, USA \\}

\author[0000-0001-5817-5944]{Blakesley Burkhart}
\affiliation{Department of Physics and Astronomy, Rutgers, The State University of New Jersey, 136 Frelinghuysen Rd, Piscataway, NJ 08854, USA \\}
\affiliation{Center for Computational Astrophysics, Flatiron Institute, 162 Fifth Avenue, New York, NY 10010, USA \\}

\author[0000-0003-3347-7094]{Piyush Sharda}
\affiliation{Leiden Observatory, Leiden University, P.O. Box 9513, NL-2300 RA Leiden, The Netherlands \\}

\author[0000-0003-2369-2911]{Claire E. Williams}
\affiliation{Department of Physics and Astronomy, UCLA, Los Angeles, CA 90095}
\affil{Mani L. Bhaumik Institute for Theoretical Physics, Department of Physics and Astronomy, UCLA, Los Angeles, CA 90095, USA\\}

\author[0000-0002-9802-9279]{Smadar Naoz}
\affiliation{Department of Physics and Astronomy, UCLA, Los Angeles, CA 90095}
\affil{Mani L. Bhaumik Institute for Theoretical Physics, Department of Physics and Astronomy, UCLA, Los Angeles, CA 90095, USA\\}

\author[0000-0001-7925-238X]{Naoki Yoshida}
\affiliation{Department of Physics, The University of Tokyo, 7-3-1 Hongo, Bunkyo, Tokyo 113-0033, Japan}
\affiliation{Kavli Institute for the Physics and Mathematics of the Universe (WPI), UTIAS, The University of Tokyo, Kashiwa, Chiba 277-8583, Japan}
\affiliation{Research Center for the Early Universe, School of Science, The University of Tokyo, 7-3-1 Hongo, Bunkyo, Tokyo 113-0033, Japan}

\author[0000-0003-3816-7028]{Federico Marinacci}
\affiliation{Department of Physics \& Astronomy ``Augusto Righi", University of Bologna, via Piero Gobetti 93/2, 40129 Bologna, Italy\\}
\affil{INAF, Astrophysics and Space Science Observatory Bologna, via Piero Gobetti 93/3, I-40129 Bologna, Italy}

\author[0000-0001-8593-7692]{Mark Vogelsberger}
\affiliation{Department of Physics and Kavli Institute for Astrophysics and Space Research, Massachusetts Institute of Technology, Cambridge, MA 02139, USA}
\affiliation{Fachbereich Physik, Philipps Universit\"at Marburg, D-35032 Marburg, Germany}

\author[0000-0002-4227-7919]{William Lake}
\affiliation{Department of Physics and Astronomy, Dartmouth College, Hanover, NH 03755, USA \\}

\begin{abstract}
We present a suite of radiation–magnetohydrodynamics simulations from the \textsc{Popsicle} project that follow the long-term growth ($\sim 50$ kyr) of primordial protostars while self-consistently coupling radiation, turbulence, and magnetic fields. The simulation suite is designed to quantify the relative impacts of the pathways of radiative feedback in Pop III stars -- the extreme-ultraviolet (EUV) ionization and Lyman-Werner (LW) dissociation -- by considering simulations with/without their inclusion. We find that without H\,\textsc{i} shielding, LW feedback can suppress and ultimately terminate accretion. With H\,\textsc{i} shielding, the large column densities near the protostar significantly weaken LW feedback. In the polar direction, atomic hydrogen fully shields LW radiation where H$_2$ self-shielding alone is insufficient. This leads to lower gas temperatures near the protostar and higher accretion rates, yielding larger final stellar masses than in models without shielding. The H\,\textsc{ii} region remain confined, extending $\sim$100 AU beyond the sink accretion radius (75 AU), as dense gravitationally bound gas sustains high recombination rates and prevents sustained pressure-driven breakout. Turbulence and magnetic fields may also contribute to its confinement, even at high ionizing luminosities. These results demonstrate that the interplay of gas dynamics, shielding, and radiative feedback can significantly alter the growth of Pop III stars. We discuss the implications for the initial mass function of primordial stars and the influence of feedback from early stellar populations.
\end{abstract}

\keywords{Star Formation, Cosmology, Population III Stars, Magnetohydrodynamical Simulations, High-redshift galaxies}

\section{Introduction}

The first stars in the Universe, also known as Population III (Pop III), form around $z \sim$ 30 from metal-free and dust-free gas that condenses at the center of dark matter (DM) minihalos with  virial masses $M_\mathrm{vir} \sim 10^{5-6}\ \mathrm{M_\odot}$ \citep[e.g.,][]{Haiman_96,Tegmark1997}. The pristine gas is initially adiabatically compressed to $n \simeq 10 \ \mathrm{cm}^{-3}$ at the center of the minihalo and subsequently cools and contracts via H$_2$ and HD ro-vibrational line emission up to $n \simeq 10^4 \ \mathrm{cm}^{-3}$ where the level populations reach local thermodynamic equilibrium, and line cooling saturates. The gas continues to contract more slowly until the enclosed mass becomes Jeans-unstable at $\simeq$ 1000 M$_\odot$ and fragments \citep[e.g.,][]{Omukai1998, Abel_00, Bromm2002, Yoshida2003,Omukai_05}. 

Over the next $\sim$ 10$^{5-6}$ years, the gas density increases quasi-isothermally by more than 15 orders of magnitude, driven by gravitational compression and thermochemical processes such as three-body H$_2$ formation \citep[e.g.,][]{Palla_83}, collision-induced emission \citep[e.g.,][]{Ripamonti_04}, and H$_2$ dissociation \citep[e.g.,][]{Omukai2010, Yoshida_08}. Following H$_2$ dissociation, the effective equation of state becomes progressively stiffer and adiabatic compression is finally halted when the pressure gradient overcomes gravity at $n \gtrsim 10^{20} \ \mathrm{cm}^{-3}$ and a hydrostatic protostar forms with a mass of $\sim$ 0.001 -- 0.01\ $\mathrm{M_\odot}$ \citep[e.g.,][]{Omukai1998, Ripamonti_02, Yoshida_08, Greif_12}. These protostars subsequently accrete gas from a protostellar disk (which often fragments) and grow to larger masses than present-day stars due to higher gas temperatures in the absence of metals and dust at zero metallicity. 

The final masses of these stars determine both their impact on the surrounding environment and their observational signatures. Assuming negligible rotation and minimal mass loss -- which primarily affect the size of the helium core and hydrogen envelope, respectively -- stars with  $M\simeq 10 - 40\ \mathrm{M_\odot}$ are expected to explode as core-collapse supernovae, those with masses $M\simeq 140 - 260\ \mathrm{M_\odot}$ as pair-instability supernovae, while those with masses $M \simeq 40 - 140\ \mathrm{M_\odot}$ or $M \gtrsim 260\ \mathrm{M_\odot}$ collapse into black holes, although uncertainties remain in these mass thresholds \citep[e.g.,][]{Heger_Woosley_02, Heger_03}. The explosion mechanism determines both the degree of metal enrichment and the extent of mechanical feedback \citep[e.g.,][]{Sluder_16,2021MNRAS.506.5247L,2025ApJ...995..165J}, and gives rise to distinct observational signatures, some of which may be detectable \citep[e.g.,][]{Tanaka_13}. Accurately predicting the final masses of Pop III stars is therefore essential for understanding how their deaths shape the early Universe.

The signatures of Pop III stars may be observable through multiple channels including the possibility of direct detections. Recent JWST candidates such as LAP1b \citep{Nakajima_25} have been proposed as potential Pop III stars at z $\sim 6.6$ consistent with key theoretical expectations \citep{2020MNRAS.497.2839L,Visbal_25, Williams_25}.\footnote{These expectations include: (1) the formation of Pop III stars in extremely low-metallicity halos with virial temperatures $T_{\rm vir}\sim10^{3}$--$10^{4}\,$K, (2) a top-heavy initial mass function, and (3) the formation of small stellar clusters containing a few $\times10^{3}\,M_\odot$ in massive Pop III stars \citep{Visbal_25}.} At even higher redshifts, stellar fluxes may be amplified by $\sim 7$–$12$ magnitudes through extreme gravitational lensing during cluster caustic transits, rendering individual Pop III stars observable with JWST’s Near Infrared Camera \citep[e.g.,][]{Rydberg_13,Windhorst_18}. Pop III stars with $M \lesssim 0.8\ \mathrm{M_\odot}$ could have survived to present day in or around the Milky Way \citep[e.g.,][]{Komiya_16, Dutta_20}. In addition, the abundance patterns of extremely metal-poor stars provide indirect constraints on the properties of their Pop III progenitors \citep[e.g.,][]{Nordlander_19,Skulad_24,Arentsen_24}. Finally, the remnants of Pop III stars may be detectable through their explosions, including pair-instability supernovae \citep[e.g.,][]{Whalen_13f}, pulsational pair-instability supernovae \citep[e.g.,][]{Whalen_14}, Type IIn supernovae, and supermassive thermonuclear explosions \citep[e.g.,][]{Whalen_13c,Whalen_13g,Whalen_13h,Johnson_13b}. 

High-resolution simulations are an important tool for understanding the formation and physical properties of Pop III stars, including their mass distribution and orbital dynamics, and thus provide a theoretical framework for interpreting future observations. Such simulations incorporate key feedback physics, including stellar radiation \citep[e.g.,][]{Stacy_16,Hosokawa_16}, magnetic fields \citep[e.g.,][]{Sharda_20,2022MNRAS.511.5042S,Saad_22,Sadanari_24,Sharda_24}, and external radiation backgrounds such as Lyman-Werner (LW) \citep[e.g.,][]{O'Shea_08,Hirano+15,vanVeenen_2025} and X-ray \citep[e.g.,][]{Park_21,Park_23}, all of which play an important role in shaping the properties of the forming stars. 

Among these feedback processes, radiation in the LW band (11.2–13.6 eV) plays a particularly central role in regulating Pop III star formation, as it destroys H$_2$, the dominant coolant required for primordial gas to condense. As a result, LW radiation suppresses star formation both within individual halos and in their surroundings \citep[e.g.,][]{Omukai_99, McKee_08, Kitayama_14, Visbal_14, Schauer_15,Schauer_17, vanVeenen_2025}. 

This LW radiation can be attenuated by large columns of H$_2$ through self-shielding, as well as by atomic hydrogen through cross-shielding. When $N_{\rm HI} \gtrsim 10^{24}\ \mathrm{cm^{-2}}$, the damping wings of the H\,\textsc{i} Lyman-series absorption lines broaden sufficiently to absorb photons in the LW band \citep[e.g.,][]{Draine_96,Wolcott-Green+_11,Glover17}. Such large columns can arise when a dissociation front driven by LW radiation produces a layer of neutral hydrogen that has not yet been ionized by the advancing ionization front \citep[e.g.,][]{Glover17}.

Several studies have explored the effects of atomic hydrogen shielding in both the inter-halo and intra-halo context. In the former, \citet{Schauer_15} found that including H\,\textsc{i} shielding in addition to H$_2$ self-shielding can reduce the LW escape fraction by more than two-thirds, provided that the source star does not fully ionize its host halo. In a follow-up work, \citet{Schauer_17} examined more massive halos ($M \sim 10^{7-8} \ \mathrm{M_\odot}$) hosting stellar clusters and found that H\,\textsc{i} shielding has little effect on the LW escape fraction. Similarly, \citet{Neyer_22} showed that the critical LW background intensity required to suppress H$_2$ cooling in atomic cooling halos ($J_{\mathrm{crit}}$) increases by $\sim$ 60 --100\% when H\,\textsc{i} shielding is included. In the intra-halo context, the assumption is often that the gas is not optically thick and the LW absorption by atomic hydrogen is estimated using a global $\sigma_{\mathrm{Lyman}} = 5.23 \times 10^{-25}$ cm$^2$ \citep[e.g.,][]{Jaura_22, Park_23}. Other studies include H\,\textsc{i} shielding  without this assumption, excising the unresolved protostellar region with a sub-sink prescription in order to follow the system for $\gtrsim$ 1 Myr \citep{Toyouchi_23}.

Given the importance of LW radiation in setting the final mass of Pop III stars, in this paper, we examine how its feedback changes when H\,\textsc{i} shielding is included in addition to H$_2$ self-shielding. For this purpose, we use a suite of unique small-scale, multi-physics simulations from the \textsc{Popsicle} (POP II/III Simulations Including Chemistry, Luminosity and Electromagnetism) project \citep{Sharda_24}. These simulations self-consistently incorporate turbulence, magnetic fields, and radiation, physical processes that are rarely combined in other small-scale Pop III star formation simulations and which must be treated simultaneously to reproduce, among other things, the observed star formation rates in present-day environments \citep[e.g.,][]{Federrath_16}. We primarily focus on how H\,\textsc{i} shielding regulates the transmission of LW flux, the thermochemical state of the gas, and the resulting stellar mass. We also explore the roles of LW radiation and EUV radiation in protostellar growth. The paper is organized as follows: In Section \S~\ref{numerical} we briefly describe the \textsc{Popsicle} simulation suite, including the initial conditions and the physical processes included in each run. In Section \S~\ref{sec:Results} we show how H\,\textsc{i} shielding modifies the accretion and protostellar growth, why the H\,\textsc{ii} region remains confined near the protostar, and investigate the origin for fragmentation in one of our runs. In Section \S~\ref{sec:discussion} we place our findings in the context of prior work, discuss the relative roles of FUV and EUV feedback, and summarize our conclusions.

\section{Methods}
\label{numerical}

\subsection{Simulation Setup}
\label{subsec:SimulationSetup}

The suite of simulations analyzed in the present study is part of the \textsc{Popsicle} project introduced in \citet{Sharda_24}. The simulation setups are similar to those employed in \citet{Sharda_25} and \citet{Sharda_24}. Here we briefly outline the main aspects of the simulation suite. We use a modified version of the magneto-hydrodynamics (MHD) adaptive mesh refinement code (AMR) FLASH \citep[][]{Fryxell_00, Dubey_08} which uses the \texttt{Paramesh} package for AMR on the grid \citep[e.g.,][]{MacNeice_00}. We solve the compressible MHD equations using an implementation of the Bouchut solver in FLASH \citep[e.g.,][]{Waagan_2011}, and use the tree-solver of \citet{Wunsch_18} to solve the Poisson equations for self-gravity. 

We model radiation feedback using the Variable Eddington Tensor–based Transport on Adaptive Meshes (\texttt{VETTAM}; \citealt{Menon_22}), which solves the radiation moment equations in an implicit formulation with closure provided by a variable Eddington tensor obtained from a time-independent, non-local ray-tracing calculation. The radiation field is divided into three energy bands: a LW band spanning 11.2–13.6 eV, corresponding to far-ultraviolet (FUV) photons that photodissociate H$_2$; an extreme-ultraviolet (EUV) band between 13.6 and 15.2 eV, which ionizes atomic hydrogen; and a higher-energy EUV band above 15.2 eV, capable of ionizing both atomic and molecular hydrogen. We do not consider the ionization of helium in this work.

The radiation is fully coupled to a primordial thermo-chemical network implemented in FLASH with the \texttt{KROME} wrapper \citep[][]{2014MNRAS.439.2386G}; the network includes the species \texttt{H, H$^+$, H$^-$, H$_2$, H$_2^+$, D, D$^+$, D$^-$, HD, HD$^+$, He, He$^+$, He$^{++}$}, and $e^-$, the primordial gas heating/cooling channels, along with heating due to photoionization, photodissociation and radiative pumping. A full description of the thermochemistry and its coupling to the radiation will be presented in a forthcoming paper (Menon et al. 2026 in preparation). 

We employ sink particles to represent individual protostars. A sink particle is inserted once gravitational collapse reaches the maximum level of AMR refinement such that the Jeans length computed from the cell density and temperature can no longer be resolved sufficiently -- i.e. the Truelove criterion \citep{Truelove_1997} -- and fulfills the additional set of criteria outlined in \citet{Federrath_10b} \footnote{The gas must also be converging, have a central gravitational potential, be Jeans unstable, be bound and not be within the accretion radius of an existing sink.}. Upon creation, sinks inherit the mass of the cell from which they form. The dynamics of the sinks are followed with a direct $N$-body second-order leapfrog integrator, and their accretion radii and softening length are set to 2.5 times the minimum cell size \citep{Federrath_10b,Federrath_2011}. The sinks act as a source term in the (non-relativistic) radiative equations and emit radiation in each band as
\begin{equation}
    \label{eq:jstar}
    j_{*}(r)=\frac{L_{*}}{\left(2 \pi \sigma_{*}^{2}\right)^{3 / 2}} \exp \left(-\frac{r^{2}}{2 \sigma_{*}^{2}}\right), 
\end{equation}
where $r$ is the radial distance of a grid cell from the sink particle and $\sigma_* = 2 \Delta x_{\mathrm{min}}$, where $\Delta x_{\mathrm{min}}$ is the minimum cell size in the domain and $L_*$ is the luminosity of the star in a given band. The luminosity is estimated from the mass and the instantaneous accretion rate of the sink particle using the modified version of the 1D \texttt{GENEVA} stellar evolution code \citep[e.g.,][]{Eggenberger_08} adopted in the stellar evolution model of \citet{Haemmerle_18}. We do not include radiative heating from accretion luminosity, which can raise gas temperatures near the protostar but does not substantially alter accretion rates and only delays but does not halt fragmentation \citep{Smith_11}. Because it is omitted across all runs, we do not expect it to affect the relative differences that are the focus of this work.

We discretize the computational domain on a base grid of resolution $64^3$, with eight additional levels of AMR, reaching a maximum spatial resolution of $\Delta x = 30\,\mathrm{AU}$ and a maximum gas density of $n_{\max} \simeq 5 \times 10^{12}\,\mathrm{cm^{-3}}$. We adopt a conservative Jeans refinement (Truelove) criterion, resolving the local Jeans length with 64 cells, significantly higher than other radiation-hydrodynamics simulations. This high Jeans resolution is essential not only for capturing small-scale dynamo amplification of magnetic fields, but also for accurately resolving shock heating. As shown by \citet[Appendix A]{Sharda_21}, insufficient Jeans resolution artificially broadens shocks, allowing gas to cool while traversing the shock front and thereby suppressing physically important shock heating. By adequately resolving the Jeans length, gas crosses shocks on timescales shorter than the cooling time, enabling a physically realistic coupling between hydrodynamics and thermochemistry.

\subsubsection{LW Radiation}
\label{subsubsec:LW Radiation}

We pause to describe some specific aspects relevant to the LW radiation from sink particles. We model the shielding of LW photons by atomic/molecular hydrogen with a sink-based ray-tracing scheme \citep{Menon_22}. We ray-trace through the adaptive mesh and integrate number densities along rays to obtain the column densities of atomic and molecular hydrogen from each radiation source. These columns are used to compute shielding factors, which attenuate the LW flux in each cell, yielding the effective LW radiation field that enters the photochemical and thermal evolution of the gas. We use the fitting functions provided by \citet{Wolcott-Green+_11} for these purposes. The cross-shielding due to H\,\textsc{i} is given by their Equation 15:
\begin{equation}
f_{\rm sh, HI} = \frac{1}{\left(1 + x\right)^{ \beta}}
\times \exp\left({\rm-\gamma}\; x\right),
\label{eq:fHIfit}
\end{equation}
 where $\beta = 1.6, ~\gamma = 0.15$ and $x = N_{\rm HI}/(2.85 \times 10^{23}\,{\rm cm^{-2}})$. The H$_2$ self-shielding factor is given by their Equation 12:
\begin{multline}
f_{\rm sh, H_2} =
\frac{0.965}{\left(1 + x/b_5\right)^\alpha}
+ \frac{0.035}{\left(1 +  x\right)^{0.5}}\\
\times \exp\left[-8.5 \times 10^{-4} \left( 1 +  x\right)^{0.5}\right],
\label{eq:fH2fit}
\end{multline}
 where $\alpha = 1.1$, $x = N_{\rm H_2}/(5 \times 10^{14}\,{\rm cm^{-2}})$ and $b_5 \equiv b/(10^{5}\,{\rm cm\,s^{-1}})$ is the normalized Doppler parameter, which we set to 1. At the characteristic H$_2$ column densities relevant for the polar regions in our runs ($N_{\rm H_2} \sim 10^{18}\,{\rm cm^{-2}}$; see Section \ref{subsec:The Impact of H Shielding}), varying  $b_5$ from our fiducial value of 1 to 0.5 or 1.5 changes $f_{\rm sh,H_2}$ by $\approx 10\%$. Small uncertainties in the adopted Doppler parameter are therefore unlikely to have a significant impact on our results. The total shielding factor is taken as the product of the contributions from H\,\textsc{i} and H$_2$ ($f_{\rm sh} = f_{\rm sh, HI} f_{\rm sh,H_2}$). \citet{Wolcott-Green+_11} showed that this separable approximation reproduces the radiative transfer calculation to within a factor of $\sim 2$ for $10^{22} \lesssim$ $N_{\rm HI}$ $\lesssim 10^{24}\,\mathrm{cm^{-2}}$. In our simulations, $N_{\rm HI} \simeq 10^{26-27}\,\mathrm{cm^{-2}}$, where overlapping Lyman-series damping wings render the LW band optically thick over the frequencies that dominate H$_2$ dissociation. In this limit H\,\textsc{i} opacity dominates the effective optical depth, and the total shielding asymptotically approaches $f_{\rm sh} \approx f_{\rm sh,HI}$.

Physically, these shielding factors represent the fraction of incident LW photons available for H$_2$ photodissociation after absorption by a band of ro-vibrational transitions of H$_2$ and electronic transitions of atomic hydrogen in the LW band. In the simulation, the total shielding factor attenuates the LW flux relative to the optically thin case; the resulting shielded flux is then used to scale the optically thin LW photodissociation rate coefficient, thereby reducing the local H$_2$ dissociation rate.

LW photons contribute to gas heating through both dissociation and excitation processes. LW photons dissociate H$_2$ via the two-step Solomon process: an H$_2$ molecule in the ground electronic state X$^{1}\Sigma_g^{+}$ absorbs a LW band photon and is excited to the B$^{1}\Sigma_u^{+}$ or C$^{1}\Pi_u$ state. Radiative decay returns the molecule to the ground electronic state, and in $\sim$10 -- 15\% of the time, the decay proceeds into the vibrational continuum rather than a bound state, resulting in dissociation into two H atoms. In the remaining non-dissociative absorptions, the excited molecule cascades through bound vibrational levels of the ground state. In dense environments ($n \gtrsim 10^4 \,\mathrm{cm^{-3}}$), collisional de-excitation efficiently converts this vibrational energy into thermal energy of the gas, depositing $\sim$ 2 eV per absorption event. In the simulation, this pumping is implemented as a local energy source term using the collisional de-excitation rate given by Equation~47 of \citet{Baczynski_15}. Photodissociation of H$_2$ also contributes to heating, with approximately 0.4 eV of kinetic energy deposited into the gas per dissociation event. These contributions are applied self-consistently as local heating terms in the energy equation.

LW radiation can also indirectly raise the gas temperature by suppressing the endothermic cooling channel associated with collisional H$_2$ dissociation. Collisional reactions such as
\begin{equation}
\mathrm{H}_2 + \mathrm{H} \rightarrow 3\mathrm{H}, \qquad
\mathrm{H}_2 + e^- \rightarrow 2\mathrm{H} + e^-,
\end{equation}
remove $\sim 4.48$ eV of thermal energy from the gas per event \citep[e.g.,][]{1979ApJS...41..555H}. In the absence of strong LW radiation, this energy loss partially balances the heating associated with H$_2$ formation via reactions such as
\begin{equation}
3\mathrm{H} \rightarrow \mathrm{H}_2 + \mathrm{H}, \qquad
\mathrm{H}^- + \mathrm{H} \rightarrow \mathrm{H}_2 + e^-.
\end{equation}
When photodissociation dominates over collisional dissociation, however, the energy required to break molecular hydrogen is supplied by the radiation field rather than the thermal reservoir. As a result, H$_2$ formation heating is no longer offset by collisional dissociation, leading to a net increase in the gas temperature.
 
\subsection{Initial Conditions}
\label{subsec:IC}

We initialize a spherical gas cloud with $M = 1000 \ \mathrm{M_\odot}$ and radius 1 pc in a Cartesian domain of side length 2.4 pc. The initial thermodynamic and chemical state of the gas is set to that obtained for a one-zone collapse evolution, similar to those in \citet{Omukai_05}, performed with our thermochemical network. This corresponds to a gas temperature of 265 K, and mass fractions $x_{\rm{H}}$ = 0.7502, $x_{\rm{H_2}}$ = 0.0006, and $x_{\rm{He}}$ = 0.2492. We also set an initial solid-body rotation around the $\hat{z}$ axis, with initial rotational energy that is 3 \% of the gravitational energy ($\Omega \sim 2 \times 10^{-14}$ rad s$^{-1}$). These conditions are meant to emulate pre-collapse conditions at the centers of minihalos at z $\sim$ 30 \citep[e.g.,][]{Abel_02, Bromm2002,yoshida+06, Hirano+14}. 

The initial turbulent velocity is $1.8\,\mathrm{km\,s^{-1}}$, corresponding to a turbulent Mach number of unity. We inject a mixture of compressive and solenoidal modes with a velocity power spectrum of $P_v \propto  k^{-1.8}$, representing a value between the Kolmogorov $k^{-5/3}$ and Burgers $k^{-2}$ turbulence \citep[e.g.,][]{Kolmogorov_41, BURGERS_1948}. We initialize the turbulence using an Ornstein-Uhlenbeck process through the method outlined in \citet{Federrath_2010} and made publicly available in \citet{Federrath_2022}. We set the initial magnetic field to 28.4$ \mu$G with a power spectrum of $P_{mag}$ $\propto k^{1.5}$ for $2 \leq k \leq 20$, representing an initial magnetic field that is amplified by the turbulent dynamo and saturates to 10$\%$ of the turbulent kinetic energy, as is expected for transonic turbulence and low Prandtl numbers at high redshift \citep[e.g.,][]{2014ApJ...797L..19F,Sharda_20}. 

We perform a suite of four simulations to investigate how H\,\textsc{i} shielding and varying radiative feedback prescriptions influence protostellar growth. All runs adopt the same turbulent seed as in \citet{Sharda_24} and \citet{Sharda_25}, but differ in the implemented radiation physics: (1) a \texttt{Fiducial} run that includes both EUV and LW radiation, with LW shielding by H$_2$ and H\,\textsc{i}; (2) a \texttt{No-Hshield} run with EUV and LW radiation, but with only H$_2$ shielding; (3) a \texttt{LW-Only} run that includes only LW radiation, with shielding by both H$_2$ and H\,\textsc{i}; and (4) a \texttt{EUV-Only} run that includes EUV radiation alone. The key differences among these simulations along with their final simulation time are summarized in Table~\ref{tab:tab1}.

\begin{figure}[htbp]
    \centering
    \includegraphics[width=\textwidth]{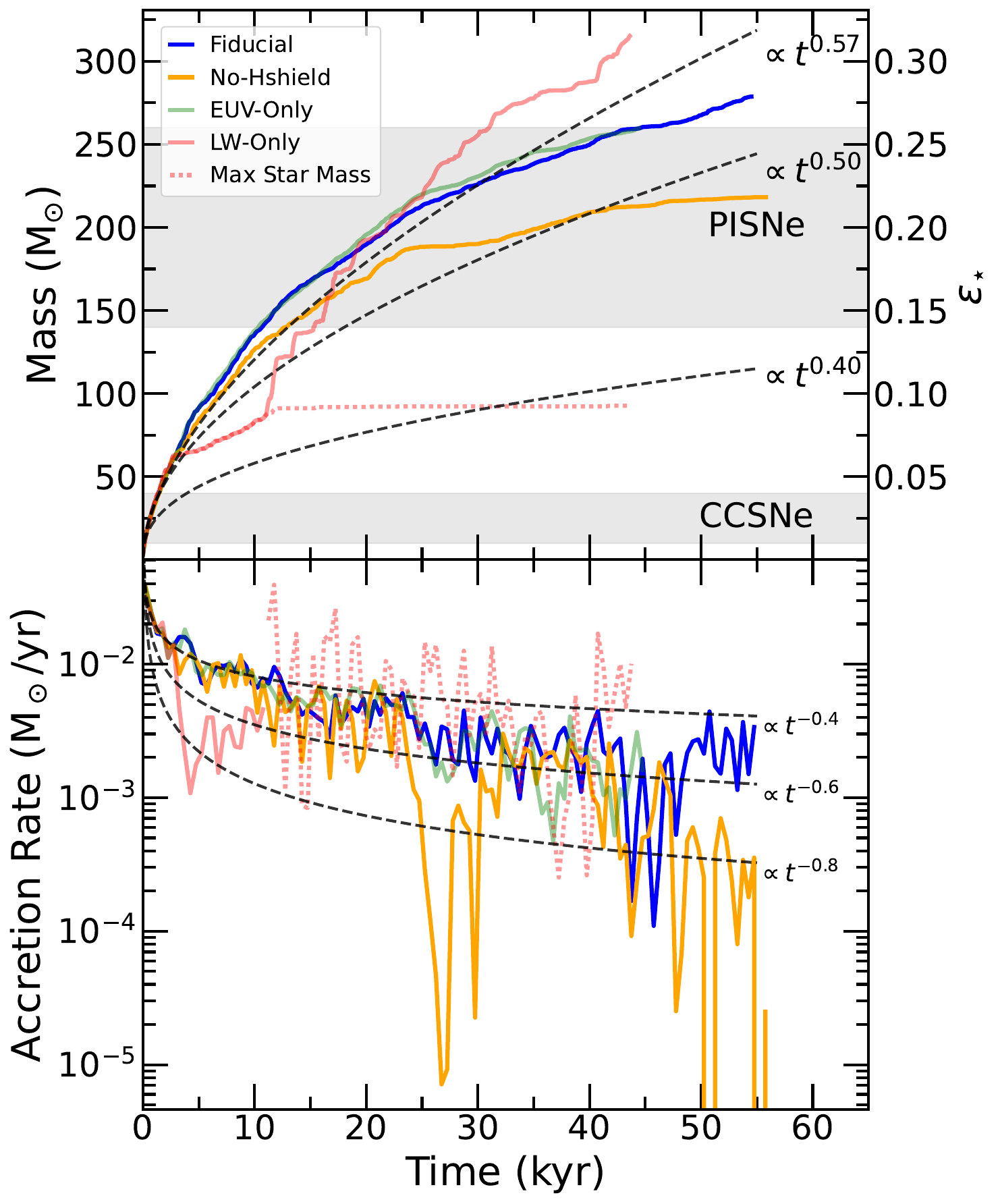}
    \caption{\textbf{Top panel:} Total stellar mass (left axis) and star formation efficiency, $\epsilon_\star$ (right axis), as a function of time since the formation of the first protostar for all four runs. In the run without atomic hydrogen shielding (No-Hshield; yellow curve), the stellar mass is suppressed relative to the Fiducial run (blue curve), which includes H\,\textsc{i} shielding. In the LW-Only run (pink curve), stellar mass growth slows markedly after formation, and the disk fragments at $\sim 11$ kyr, ultimately producing a total of 11 sink particles by the end of the simulation. The dashed pink curve shows the mass of the most massive sink. Dashed black lines indicate power-law scalings for reference. Grey-shaded bands denote the stellar mass regime in which Pop III stars are expected to explode as core-collapse supernovae (CCSNe) or pair-instability supernovae (PISNe). The stellar mass evolution in the EUV-Only run and the Fiducial run is similar, reflecting the effective shielding of LW radiation in the latter.
    \textbf{Bottom panel:} Total stellar accretion rate as a function of time for all four runs, averaged over $\Delta t = 500$ yr. The dashed pink curve indicates accretion onto all stars in the LW-Only run after fragmentation.}
    \label{fig:mass_accretion}
\end{figure}

\begin{table}[H]
\centering
\caption{Description of radiation and shielding implementations in each run, along with the final simulation time relative to first star formation.}
\begin{tabular}{lccccc}
\hline
Name & EUV & LW & H$_2$-Sh. & H\,\textsc{i}-Sh. & $t_{\rm end}$ (kyr) \\
\hline
\texttt{Fiducial}   & \checkmark & \checkmark & \checkmark & \checkmark & 55 \\
\texttt{No-Hshield} & \checkmark & \checkmark & \checkmark & $-$        & 56 \\
\texttt{LW-Only}    & $-$        & \checkmark & \checkmark & \checkmark & 44 \\
\texttt{EUV-Only}   & \checkmark & $-$        & $-$        & $-$        & 45 \\
\hline
\end{tabular}
\label{tab:tab1}
\end{table}

\section{Results}
\label{sec:Results}

\subsection{General Evolution}
\label{subsec:General Trends}

Across all the runs, the gas cloud undergoes nonhomologous runaway gravitational collapse, characterized by a monotonic increase in central density with a characteristic density profile that varies outside the core as $r^{-2.3}$ \citep[e.g.,][]{Omukai1998, Ripamonti_02}. The first sink, which we hereafter refer to as the protostar unless otherwise noted, forms at 1.5 Myr, corresponding to roughly 2.8 times the free-fall time at the initial density, owing in part to the combined support from rotation and magnetic fields. A longer collapse time can facilitate HD formation and allow HD cooling to contribute to cooling the gas \citep[e.g.,][]{Hirano+14}. In the snapshot prior to sink formation, we find $133\ \mathrm{M_\odot}$ of gas cooler than $300$ K, but only $3\ \mathrm{M_\odot}$ below $200$ K, with a minimum temperature of $158$ K. Since  HD cooling typically cools primordial gas down to $\sim 50$ - $100$ K, it is unlikely to be the dominant coolant in our simulations \citep[e.g.,][]{Ripamonti_07}. The sink forms with a mass of $\sim 3\ \mathrm{M_\odot}$. This relatively large initial sink mass reflects the resolution scale of the simulation and may slightly advance the onset of radiative feedback. Since the runs begin to diverge only after the stellar mass exceeds $\sim 60\ \mathrm{M_\odot}$ (see Figure~\ref{fig:mass_accretion}), we do not expect the initial sink mass to materially affect the relative accretion histories.

\begin{figure}[htbp]
    \centering
    \includegraphics[width=\columnwidth]{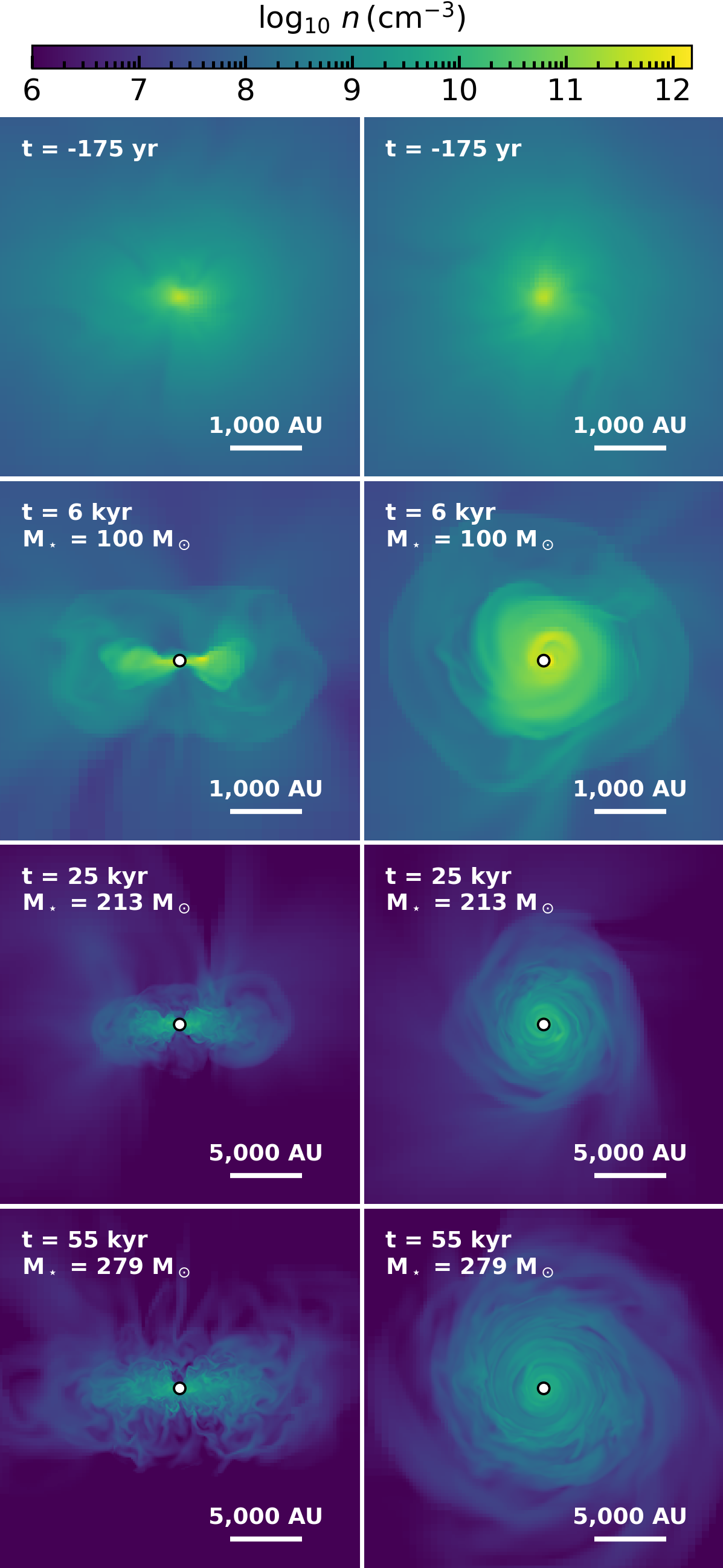}
    \caption{ Each row shows two orthogonal slice plots of gas number density at
a different evolutionary stage in the Fiducial run. The top row shows the dense, oblate structure that forms at the center of the collapsing gas cloud immediately prior to protostar formation. The second row shows the emergence of a flattened, rotationally supported disk a few thousand years after collapse. The third and fourth rows show the subsequent growth of the disk and surrounding envelope as the system accretes mass over time. The white markers in the lower three rows indicate the locations of sink particles. The spatial scale increases by a factor of five between the upper and lower rows, as indicated by the scale bars.}
    \label{fig:density_slices}
\end{figure}

In Figure~\ref{fig:density_slices} we show orthogonal slices of the gas number density at four different epochs centered on the densest gas in the first row and the sink particle in the subsequent rows. The first row shows the densest gas at t $\sim$ 175 yrs prior to the creation of the protostar, where a slightly oblate spheroid has formed that is about to collapse. In the second row we see a rotationally supported, protostellar disk, created by the initial angular momentum of the cloud. This disk continuously grows in mass and radial extent as it accretes gas from the envelope, and 20,000 years later, has grown in radius by a factor of around 5, with a typical temperature of around 400 K in the disk. In the final snapshot (bottom row), the disk and central protostar continue to grow in mass and size as accretion has not yet been terminated.

In all runs except the LW-Only case, a single protostar forms and grows to a distinct final mass by the end of the simulation. The final stellar masses are $279\ \mathrm{M_\odot}$ in the Fiducial run, $218\ \mathrm{M_\odot}$ in the No-Hshield run, and $260\ \mathrm{M_\odot}$ in the EUV-Only run, while the most massive star in the LW-Only run reaches $93\ \mathrm{M_\odot}$ by the end of the simulation. The top panel of Figure~\ref{fig:mass_accretion} shows the evolution of the total stellar mass as a function of time since sink formation for all four runs, with the corresponding star formation efficiency indicated on the right-hand axis. During the first $\sim$ 3~kyr, the Fiducial and No-Hshield runs exhibit comparable mass growth. Thereafter, the stellar mass in the Fiducial run increases more rapidly, owing to the shielding of LW radiation by H\,\textsc{i}. As a result, the mass growth history of the Fiducial run closely resembles that of the EUV-Only run, reflecting that the final stellar mass in each run is set primarily by the implemented radiative feedback.

The bottom panel of Figure~\ref{fig:mass_accretion} shows the accretion rate onto the sink particle, averaged over 500~yr intervals, for all four runs. In the Fiducial run, the accretion rate initially declines but subsequently stabilizes from  $\sim$ 25 kyr onward, aside from a transient dip near $\sim$ 45 kyr. In contrast, the No-Hshield run exhibits a sharp suppression of accretion driven by LW heating at $\sim$ 25 kyr. This is followed by a second, more pronounced decline near $\sim$ 50~kyr that effectively terminates further accretion. 

In contrast, the LW-Only run fragments and produces the largest total stellar mass among all runs by the end of its run. As shown in Figure~\ref{fig:mass_accretion}, the accretion onto the central protostar slows markedly at $\sim$ 3 kyr and the disk fragments $\sim$ 11 kyr after its formation. Following an initial fragmentation, the disk continues to fragment, ultimately producing a total of 11 sink particles by the end of the simulation. The enhanced total stellar mass likely results from weak effective feedback: EUV radiation is absent, and H\,\textsc{i} shielding attenuates the LW flux. We analyze the underlying physical mechanisms responsible for the fragmentation in Section \ref{subsec:Frag_FUV}. 

\subsection{The Impact of Atomic Hydrogen Shielding}
\label{subsec:The Impact of H Shielding}

\begin{figure}[htbp]
    \centering
    \includegraphics[width=\textwidth]{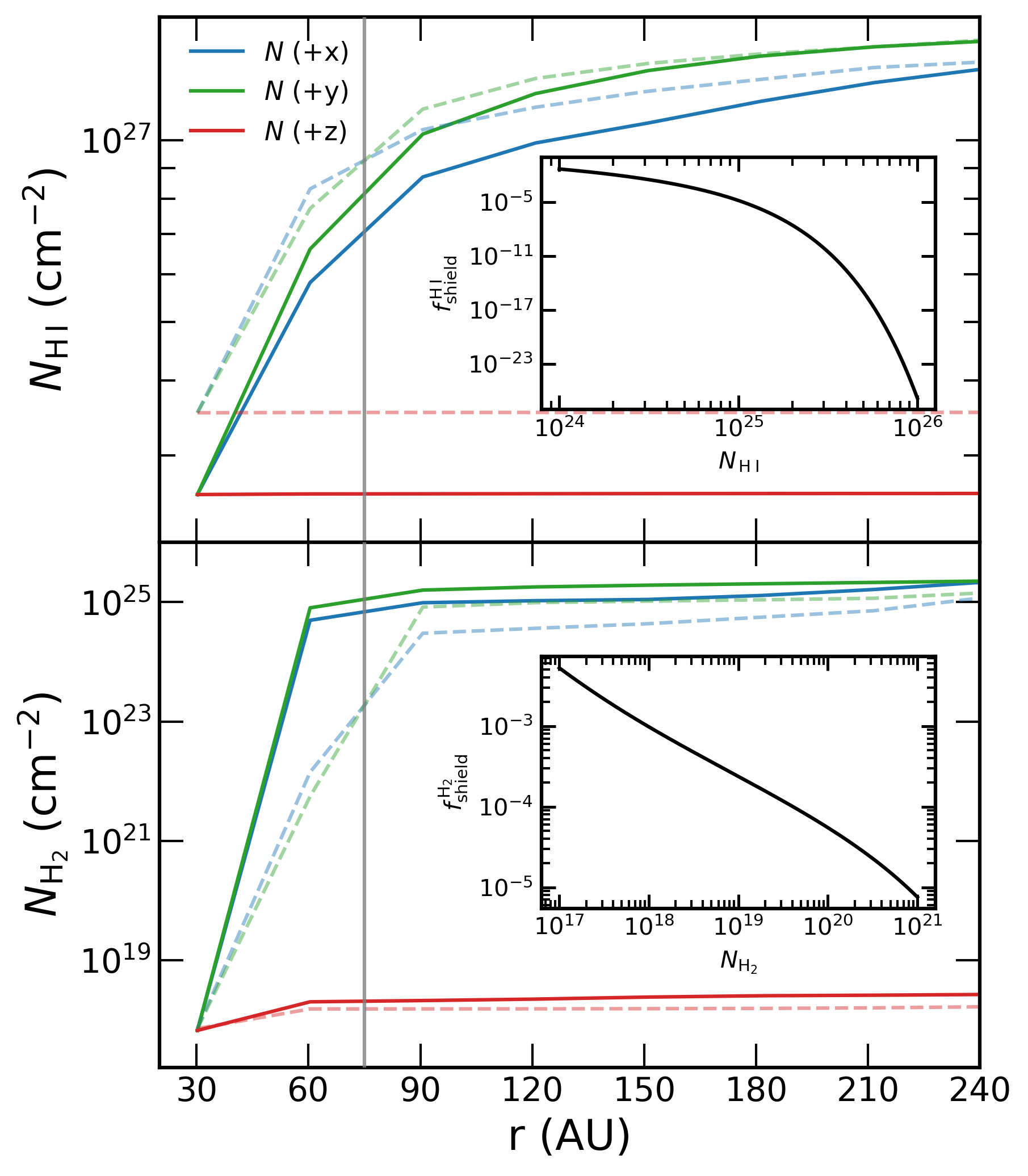}
    \caption{
    Column densities of atomic hydrogen, $N_{\rm HI}$ (top panel), and molecular hydrogen, $N_{\mathrm{H_2}}$ (bottom panel), measured from the protostar along the positive $x$ (blue), $y$ (green), and $z$ (red) directions for the Fiducial run (solid lines) and No-Hshield (dashed lines) at t $\simeq$ 6 kyr and M $\simeq 100\ \mathrm{M_\odot}$, corresponding to the same snapshot as the second row of Figure~\ref{fig:density_slices}. The $z$-axis corresponds to the polar direction, while $x$ and $y$ are radial directions in the disk plane. Insets show the corresponding shielding functions, $f_{\rm sh, HI}$ and $f_{\rm sh, H_2}$ from Equations \ref{eq:fHIfit} and \ref{eq:fH2fit} for each species. The gray vertical line marks the sink accretion radius, $r_{\rm sink}=75$ AU.}
    \label{fig:column}
\end{figure}

\begin{figure}[htbp]
    \centering    
    \includegraphics[width=\textwidth]{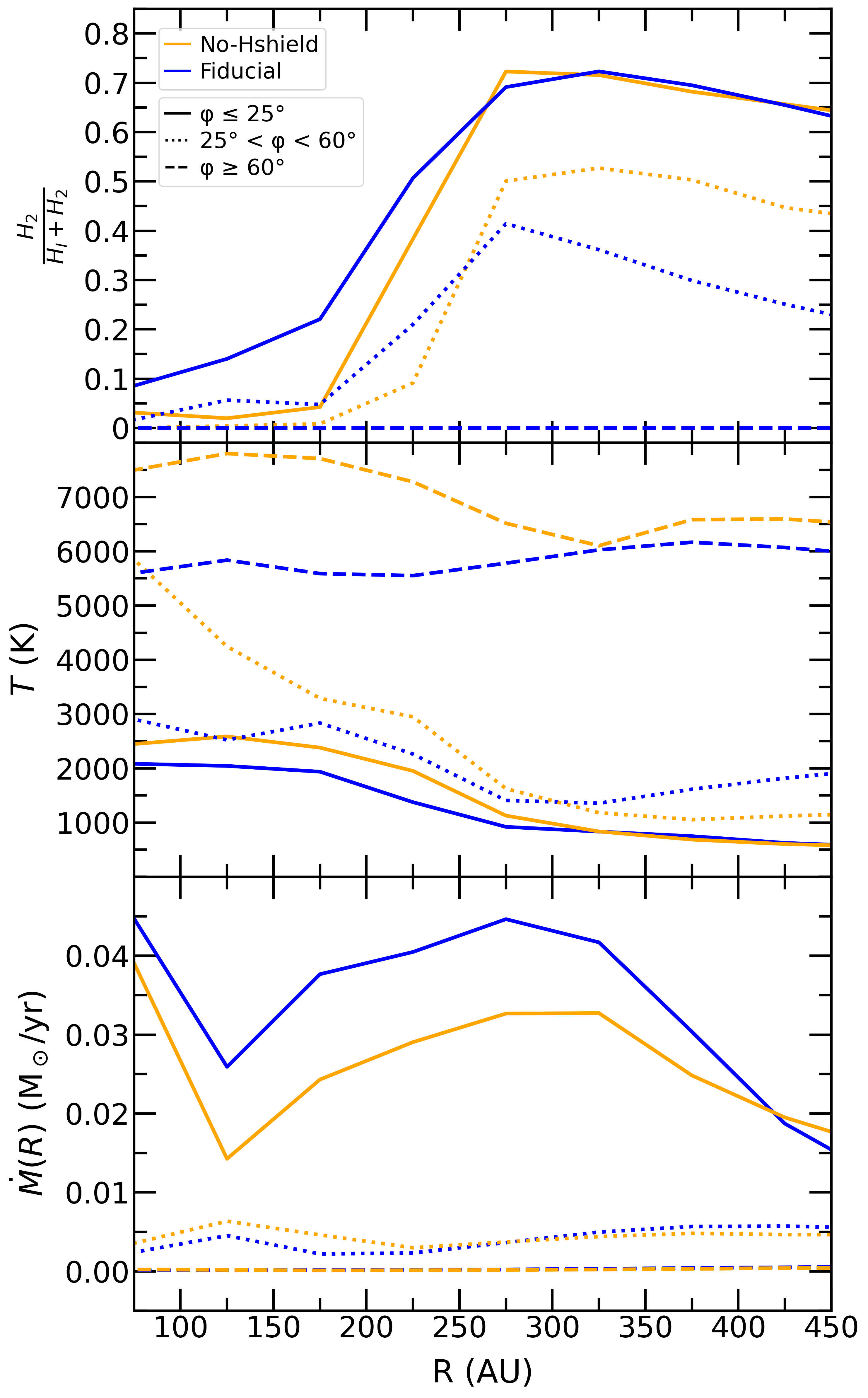}
    \caption{Mass-weighted radial profiles for the Fiducial (blue lines) and No-Hshield (yellow lines) runs, centered on the protostar at $t \simeq 6$ kyr and $M_\star \simeq 100\ \mathrm{M_\odot}$, corresponding to the same snapshot as the second row of Figure~\ref{fig:density_slices}. Profiles are grouped into three angular regions measured by the angle $\phi$ from the disk midplane: disk-dominated ($\phi \leq 25^\circ$, solid), intermediate ($25^\circ < \phi < 60^\circ$, dotted), and polar ($\phi \geq 60^\circ$, dashed).} The top panel shows the mass-weighted H$_2$ mass fraction, which is higher in the Fiducial run near the sink due to shielding of LW radiation by atomic hydrogen. The middle panel shows the corresponding mass-weighted gas temperature, which is lower in the Fiducial run as a result of more efficient H$_2$ cooling. The bottom panel shows the radial mass flow rate, which is reduced in the No-Hshield run due to enhanced thermal pressure support from the hotter gas.
    \label{fig:Spherical}
\end{figure}

The larger stellar mass in the Fiducial run relative to the No-Hshield run (Figure~\ref{fig:mass_accretion}) arises from the inclusion of H\,\textsc{i} shielding. H\,\textsc{i} shielding attenuates the incident LW radiation field, reducing H$_2$ photodissociation and preserving higher molecular abundances near the protostar. The enhanced H$_2$ fraction strengthens molecular cooling and lowers the gas temperature. This reduces thermal pressure support and facilitates more efficient inward mass transport that sustains a higher accretion rate onto the protostar. In contrast, when H\,\textsc{i} shielding is absent, stronger LW irradiation suppresses H$_2$ formation, elevates the gas temperature, increases pressure support, and ultimately limits protostellar growth. We now quantify each step of this process.

We begin by examining the cumulative column densities of atomic and molecular hydrogen in both runs shortly after they diverge when $M_* \simeq 100\ \mathrm{M_\odot}$. The column density of a species $X$ is defined as
\begin{equation}
N_X = \int n_X \, dl ,
\end{equation}
where $n_X$ is the number density of species $X$ and $dl$ is the differential path length element along the chosen pencil-beam direction. We compute this integral along a one-cell-wide beam, cast from the sink center in the positive $\hat{x}$ and $\hat{y}$ directions (corresponding to the disk plane) and in the positive $\hat{z}$ direction (corresponding to the polar direction). In the top panel of Figure~\ref{fig:column}, we show the H\,\textsc{i} column density when $M_* \simeq 100\ \mathrm{M_\odot}$ in the No-Hshield (dashed lines) and Fiducial (solid lines) run, with the inset displaying the corresponding shielding factor of Equation \ref{eq:fHIfit}. In both cases, the radial and polar directions exhibit sufficiently large column densities ($N_{\rm HI}  \simeq 10^{26-27} \mathrm{cm}^{-2}$) that can completely shield the incident LW flux, as indicated by the inset showing full shielding. 

The high H\,\textsc{i} column is set by a combination of factors and persists for the runs that do not fragment. Prior to sink formation the central gas has high density ($5.45  \times 10^{-13} \rm g\,cm^{-3}$) and a noticeable atomic fraction (0.31). Immediately after sink formation, the H$_2$ in the inner region is photodissociated, raising the H\,\textsc{i} column by an order of magnitude, similar to values seen in Figure~\ref{fig:column}. For the runs that do not fragment, this same H\,\textsc{i} column persists for the duration of the simulations and does not significantly evolve with time.\footnote{For the fragmenting LW-Only run, the H\,\textsc{i} columns surrounding some of the newly formed sinks are not high enough to fully shield the LW radiation. The dynamics of a fragmenting protostellar disk reduces the densities surrounding newly formed sinks.} We note, however, that the shielding is dominated by gas within the sink-accretion radius. In Section~\ref{subsec:resolution} we discuss how this might be affected by our adopted numerical resolution ($\sim 30 \, \rm AU$).

\begin{figure*}
  \centering
  \includegraphics[width=\linewidth]{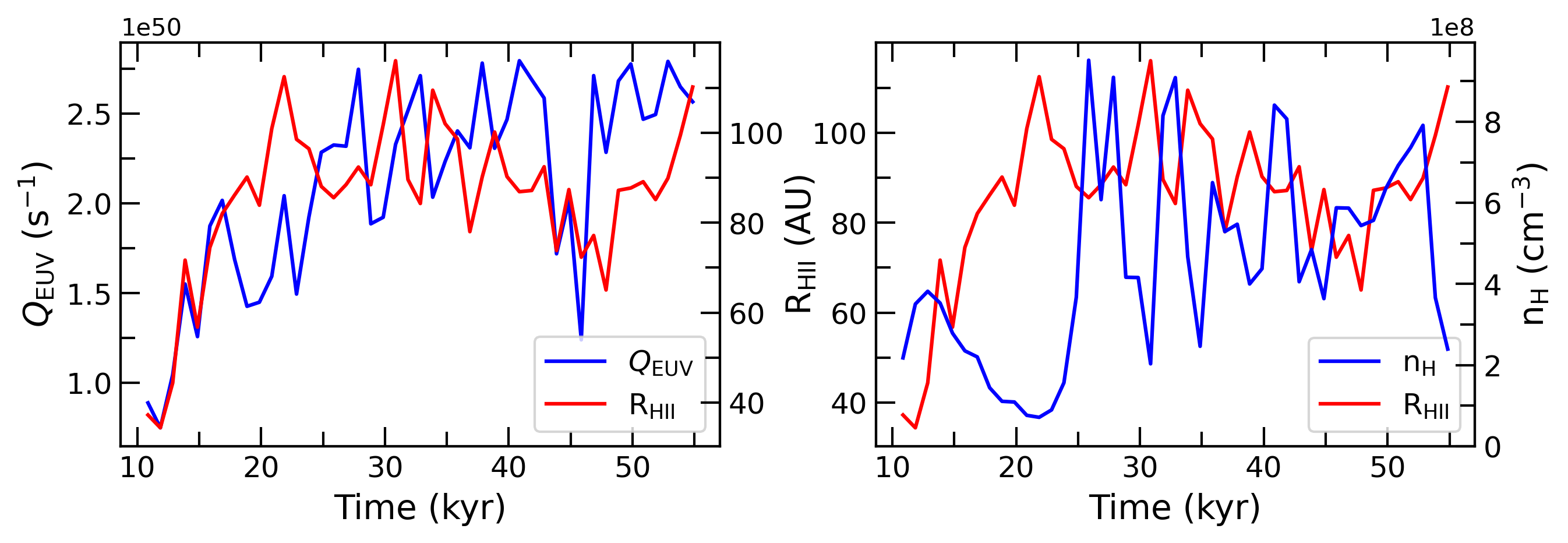}
    \caption{Evolution of the H\,\textsc{ii} region in the Fiducial run. \textbf{Left panel:} The size of the H\,\textsc{ii} region, $R_{\text{H\,\textsc{ii}}}$ (right axis), shown together with the ionizing photon emission rate, $Q_{\rm{EUV}}$ (left axis), as a function of time. \textbf{Right panel:} The H\,\textsc{ii} region size, $R_{\text{H\,\textsc{ii}}}$ (left axis), shown together with the median gas number density, $n_{\mathrm{H}}$ (right axis), measured within a cylindrical region centered on the sink particle, with a base radius of 125~AU and extending 100~AU above and below the sink accretion radius. The H\,\textsc{ii} region initially grows in tandem with the increasing EUV photon emission rate, and subsequently fluctuates in size in response to variations in both the local gas density and the ionizing luminosity.}   
   \label{fig:HII}
\end{figure*}

The bottom panel of Figure~\ref{fig:column} shows the H$_2$ column density at the same snapshot, with insets indicating the corresponding H$_2$ self-shielding factor computed using Equation \ref{eq:fH2fit}. In the radial directions ($\hat{x}$ and $\hat{y}$), corresponding to the disk plane, the H$_2$ column density remains high in both runs, reaching  $\simeq 10^{25}\ \mathrm{cm^{-2}}$, and is therefore sufficient to fully shield the gas from LW radiation. In contrast, along the polar direction ($\hat{z}$), the H$_2$ column density is lower in both runs by approximately seven orders of magnitude and by itself is insufficient to completely shield the LW radiation.

In the Fiducial run, H\,\textsc{i} provides an additional source of opacity that prevents LW radiation from breaking out along low H$_2$ columns in the polar direction. At this time the stellar LW luminosity is $4.1 \times 10^{38}\ \mathrm{erg\ s^{-1}}$ and the polar H$_2$ column density is $2.4 \times 10^{18}\ \mathrm{cm^{-2}}$. The corresponding self-shielding factor allows a transmitted LW flux capable of dissociating H$_2$, yielding, for instance, $F_{\mathrm{LW}} = 3.6 \times 10^{3}\ \mathrm{erg\ s^{-1}\ cm^{-2}}$ ($ \sim 2 \times 10^{6}$ $G_0$) at $r = 150\ \mathrm{AU}$. However, because H\,\textsc{i} shielding is included in the Fiducial run this flux is effectively shielded by the opacity of atomic hydrogen. 

By contrast, the No-Hshield run has a comparable stellar LW luminosity, $L_{\rm{LW}} = 6.9 \times 10^{38}\ \mathrm{erg\ s^{-1}}$, and a similarly low H$_2$ column, $N_{\mathrm{H_2}} = 1.5 \times 10^{18}\ \mathrm{cm^{-2}}$ in the polar direction. In the absence of H\,\textsc{i} shielding, the transmitted LW flux remains high enough to dissociate H$_2$, reaching, for instance, $ F_{\mathrm{LW}} = 8.2 \times 10^{3}\ \mathrm{erg\ s^{-1}\ cm^{-2}}$ at $r = 150\ \mathrm{AU}$. These results show that H\,\textsc{i} shielding provides a crucial additional source of opacity along the polar direction, where H$_2$ self-shielding alone is insufficient to prevent LW escape in an otherwise anisotropic radiation field.

The absence of H\,\textsc{i} shielding of LW radiation suppresses protostellar mass growth by allowing a stronger LW radiation field to penetrate the gas surrounding the sink. This raises gas temperatures in the vicinity of the protostar and enhances thermal pressure support. The enhanced thermal pressure counteracts gravity, inhibiting inward mass transport and thereby reducing the protostar's accretion rate.

This effect is illustrated in Figure~\ref{fig:Spherical}, which shows mass-weighted radial profiles of the H$_2$ mass fraction, temperature, and radial mass flux at the same snapshot as Figure~\ref{fig:column} (i.e.\ $M_* \simeq 100\ \mathrm{M_\odot}$), separated into three angular regions measured relative to the disk midplane: disk-dominated, intermediate, and polar. The top panel shows that the No-Hshield run exhibits lower H$_2$ abundances near the protostar in both the disk and intermediate regions (with negligible amounts in both runs in the polar direction), consistent with enhanced LW-driven photodissociation. In contrast, the Fiducial run retains higher molecular fractions near the sink owing to effective LW shielding. The reduced H$_2$ abundance in the No-Hshield run, together with additional LW heating mechanisms discussed in Section \ref{subsubsec:LW Radiation}, suppresses cooling and leads to higher gas temperatures over most of the plotted range as seen in the second panel. These elevated temperatures increase thermal pressure support, suppress inward mass transport, and produce a systematically reduced radial mass flux, which, as seen in the bottom panel, is dominated by gas in the disk. The resulting suppression of mass inflow limits the protostellar growth in the No-Hshield run.

\subsection{Confinement of Ionized Region}
\label{subsec:HII}

\begin{figure*}[t]
    \centering
    \includegraphics[width=\textwidth]{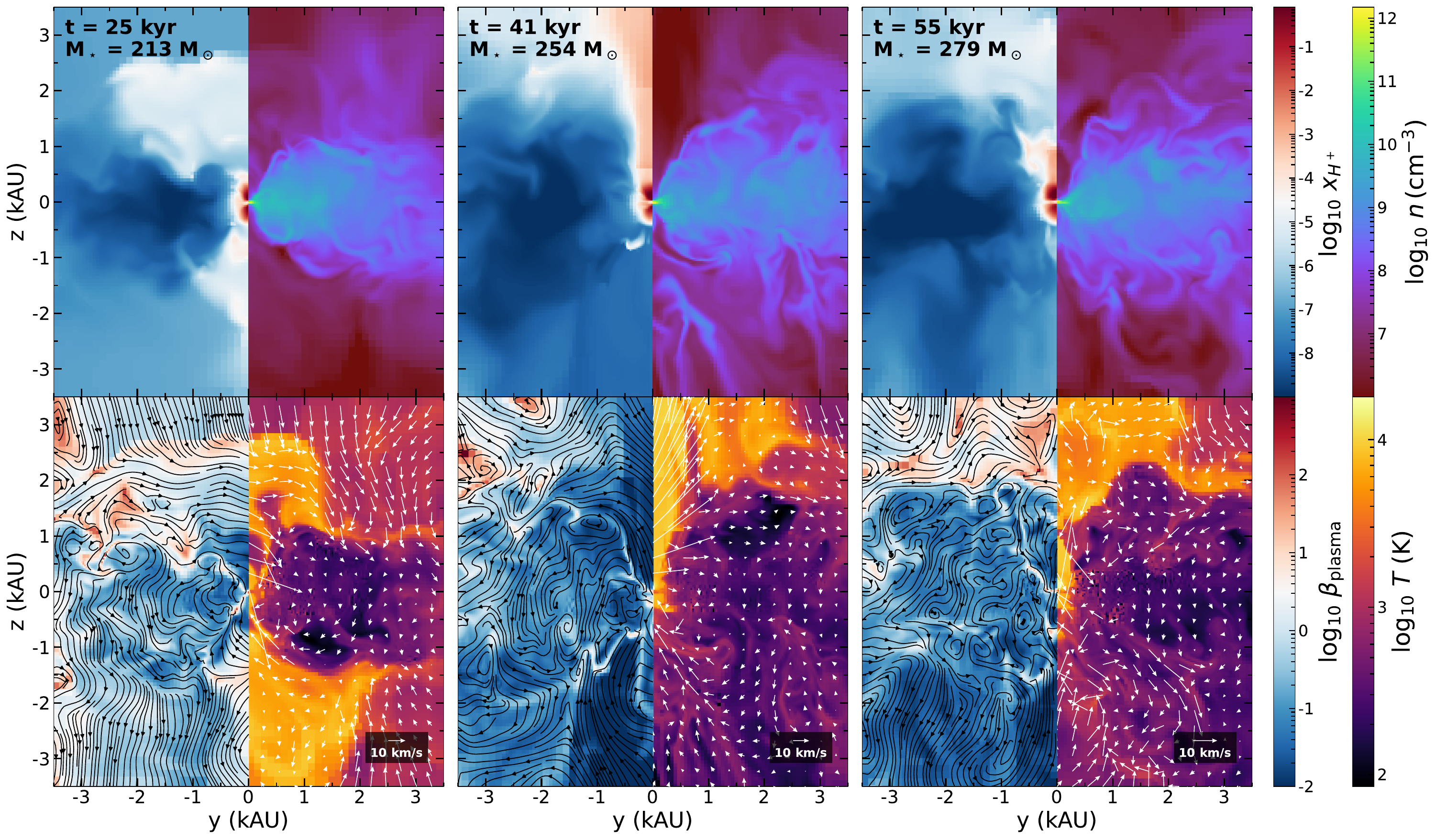}
     \caption{Multiphase structure of the gas in the Fiducial run at three evolutionary stages (t = 25, 41, and 55 kyr; left to right). \textbf{Top row:} mass fraction of ionized hydrogen, $x_{\mathrm{H^+}}$ (left half of each panel), and gas number density (right half). \textbf{Bottom row:} plasma beta, $\beta$ (left half), and gas temperature (right half). All panels show slices centered on the protostar and taken perpendicular to the simulation $\hat{x}$-axis; the left and right halves of each panel correspond to opposite sides of the protostar rather than mirror images of the same region. Magnetic field orientations are indicated by streamlines in the bottom-left panels, while gas velocity vectors are shown in the bottom-right panels. Across all times shown, the H\,\textsc{ii} region remains compact and confined to the immediate vicinity of the protostar, with short-lived, intermittent outflows of partially ionized gas as shown in the middle panel of the top row. The snapshots shown in the first and third columns correspond to the same evolutionary stages as the third and fourth rows of Figure~\ref{fig:density_slices}, respectively.}
    \label{fig:fields_slice}  
\end{figure*}

H\,\textsc{ii} regions first emerge at t $\simeq$ 10 kyr after the formation of the protostar in all runs that include EUV radiation. In these runs, the ionized regions remain mostly confined to within R $\lesssim$ 100 AU measured outward from the sink accretion radius (75 AU). To understand the source of the confinement, we first compute the size of the H\,\textsc{ii} region in our Fiducial run. We identify cells as ionized if their hydrogen ionization number fraction satisfies $X_{\mathrm{H^+}} > 0.5$. We then sum the volumes of all these cells and define an effective radius of the ionized region as

\begin{equation}
R_{\rm HII} = \left(\frac{3}{4\pi}\sum_i V_i\right)^{1/3},
\label{eq:HII}
\end{equation}
 where $V_i$ is the volume of each ionized cell. 

In the left panel of Figure~\ref{fig:HII}, we show the EUV photon emission rate, $Q_{\rm EUV}$, alongside the size of the H\,\textsc{ii} region. The ionized region first appears at $t \simeq 10$ kyr and initially expands in tandem with $Q_{\rm EUV}$ until $t \simeq 20$ kyr. At that time the H\,\textsc{ii} region continues to grow despite a decline in $Q_{\rm EUV}$. As shown in the right panel, this coincides with a decrease in the local hydrogen number density. The plotted density corresponds to the median value measured within a cylindrical region centered on the protostar, with a base radius of 125 AU and extending 100~AU above and below the sink accretion radius. This selection targets polar gas in the ionized lobes while excluding material inside the sink accretion radius. We confirm that doubling the base or height of the cylinder does not meaningfully change the shape of the curve. At later times, the effective radius of the H\,\textsc{ii} region fluctuates between $R \sim 80 - 120$ AU above the sink accretion radius. The outer extent of the H\,\textsc{ii} region is therefore regulated by the unresolved inner disk and the smallest resolved scales (see Appendix~\ref{app:photon_consumption}).

The characteristic size of an H\,\textsc{ii} region may also be estimated analytically using the Str\"omgren radius, defined as
\begin{equation}
    R_{\mathrm{St}} = \left( \frac{3 \, Q_{\rm{EUV}}}{4 \pi \, \alpha_{\mathrm{B}} \, n_{\mathrm{H}}^2} \right)^{1/3},
    \label{eq:strom}
\end{equation}
where $Q_{\rm{EUV}}$ is the ionizing photon emission rate and $n_{\mathrm{H}}$ is the hydrogen number density. Adopting representative values in the vicinity of the protostar from Figure~\ref{fig:HII} of $n_{\mathrm{H}} \simeq 6 \times 10^{8}\ \mathrm{cm^{-3}}$ and $Q_{\rm{EUV}} \simeq 2.25 \times 10^{50}\ \mathrm{s^{-1}}$, Equation \ref{eq:strom} yields a Str\"omgren radius of $R_{\mathrm{St}} \simeq 55$ AU broadly consistent with the effective radius of the ionized region, $R_{\mathrm{H\,\textsc{ii}}} \simeq 90$ AU, measured directly from Equation \ref{eq:HII} and shown in Figure~\ref{fig:HII}. The Str\"omgren radius formally assumes ionization equilibrium in a static medium. In our simulations, this assumption is satisfied, as the recombination time in the ionized gas ($t_{\rm rec}\sim$ days) is orders of magnitude shorter than both the hydrodynamic timestep and local dynamical timescales. The effective radius of the H\,\textsc{ii} region exhibits no sustained expansion, implying an ionization-front speed $v_{\rm IF}\ll c_{\rm{s}}$. The ionized region therefore resides in a quasi-static, density-confined regime rather than undergoing a propagating breakout.

An H\,\textsc{ii} region remains gravitationally trapped as long as the ionized gas is bound to the central protostar. A useful dynamical scale is therefore the gravitational radius, defined as the radius where the ionized-gas sound speed is comparable to the local Keplerian velocity -- equivalently, where the thermal energy scale per unit mass of the ionized gas is comparable to the gravitational potential scale -- giving
\begin{equation}
    r_g \sim \frac{GM_*}{c_{s,{\rm HII}}^2},
\end{equation}
up to factors of order unity \citep[e.g.,][]{Liffman_03, McKee_08, Tanaka_13}. In our simulations the radius of the H\,\textsc{ii} region never exceeds this value. For instance, at t = 25 kyr and t = 55 kyr (same time as first and last panel in Figure~\ref{fig:fields_slice}), the stellar masses are $213\ \mathrm{M_\odot}$ and $279\ \mathrm{M_\odot}$ while the mass-weighted ionized gas sound speeds are $12.8$ and $16.7\,{\rm km\,s^{-1}}$, giving gravitational radii of $r_g=1153$ and $888$ AU, exceeding the maximum size of the H\,\textsc{ii} region. The gravitational field therefore helps prevent the expansion of the H\,\textsc{ii} region.

The large-scale density field near the protostar may also contribute to the confinement. This is illustrated in the top row of Figure~\ref{fig:fields_slice}, which shows side-by-side slices of ionized fraction (left) and gas number density (right) at three different times, the first and last of which correspond to the bottom two rows of Figure~\ref{fig:density_slices}. Each panel shows a slice centered on the protostar and taken perpendicular to the simulation $\hat{x}$-axis, with the left and right halves corresponding to opposite sides of the protostar. For the ionized region to expand and undergo a large-scale breakout, the ionized gas must become sufficiently over-pressurized to drive a sustained, pressure-driven expansion into the surrounding neutral medium. In our simulations, the strong pressure gradients required for such sustained expansion do not develop, preventing the formation of a shock-bounded, pressure-driven ionization front. Instead, the surrounding gas forms an inhomogeneous, dynamically maintained medium with striated, puffed-up dense structures extending to $\sim$ 2,000 -- 3,000 AU above and below the disk plane. Although bursts of partially ionized gas, produced by the thermal expansion of locally over-pressurized pockets, are visible in the middle panel, they are transient and drive temporary mass loss from the immediate vicinity of the protostar. 

\begin{table}[t]
\centering
\caption{Median values of thermal and magnetic pressures for selected plasma $\beta$.}
\begin{tabular}{lccc}
\hline
Time (kyr) & $\log_{10}\beta$ & $\log_{10}(P_{\rm th}/k_B)$ & $\log_{10}(P_{\rm B}/k_B)$ \\
\hline
25 & $-1$ & 11.14 & 12.14 \\
25 & $+1$ & 10.66 & 9.65 \\
\hline
41 & $-1$ & 10.40 & 11.40 \\
41 & $+1$ & 11.06 & 10.06 \\
\hline
55 & $-1$ & 10.58 & 11.58 \\
55 & $+1$ & 10.58 & 9.58 \\
\hline
\end{tabular}
\label{tab:tab2}
\end{table}

Potentially dynamically important magnetic fields surround the protostar and disk. In the bottom row of Figure~\ref{fig:fields_slice} we show the plasma beta with magnetic field lines (left) and the temperature with velocity quivers (right). We see tangled magnetic fields with plasma $\beta = P_{\rm th}/P_{\rm{B}} = (n k_B T)/(B^2/8\pi) < 1$ surrounding the protostellar disk and H\,\textsc{ii} region, implying that magnetic pressure is comparable to or exceeds the thermal pressure and may therefore influence the expansion of the H\,\textsc{ii} region. For reference, in Table~\ref{tab:tab2} we show the median values of $P_{\rm th}$ and $P_B$ for $\log_{10}\beta = \pm 1$ for each slice in Figure~\ref{fig:fields_slice}. 

\begin{figure}[t]
    \centering
    \includegraphics[width=\textwidth]{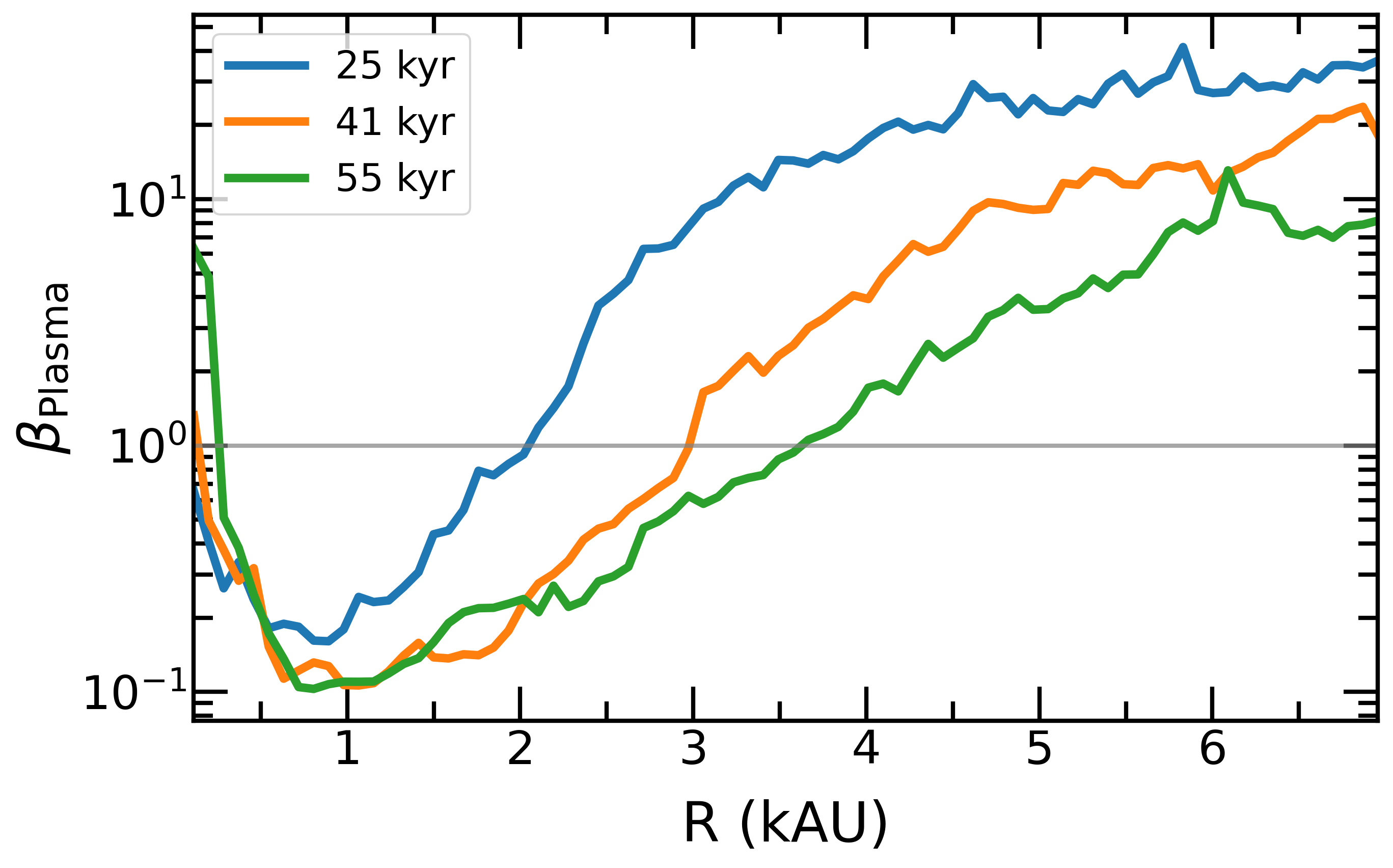}
    \caption{Mass-weighted plasma beta in the Fiducial run for the same three snapshots shown in Figure~\ref{fig:fields_slice} ($t = 25, 41,$ and $55$ kyr). The profiles show that the gas surrounding the protostellar disk and compact H\,\textsc{ii} region remains largely magnetically supported ($\beta \lesssim 1$) over a broad range of radii, with a transition toward $\beta \gtrsim 1$ at larger distances from the protostar that increases with time.}   
    \label{fig:radial_beta}
\end{figure}

To illustrate how the relative importance of thermal and magnetic pressure varies with distance from the protostar, we compute mass-weighted radial profiles of plasma $\beta$ for the same three snapshots shown in Figure \ref{fig:fields_slice}. Figure \ref{fig:radial_beta} shows that the transition from $\beta < 1$ to $\beta >1$ occurs between 2 -- 4  kAU from the sink. The radius at which $\beta = 1$ also shifts outward with time, indicating that the magnetically dominated region expands as the system evolves. 

\subsection{Fragmentation of the LW-Only Run}
\label{subsec:Frag_FUV}

As illustrated in Figure~\ref{fig:mass_accretion}, the protostellar evolution in the LW-Only run differs markedly from the other runs, forming multiple protostars rather than a single, massive one. This is contrary to expectations that LW radiation inhibits fragmentation. In our LW-Only run, however, LW radiation is effectively shielded by atomic hydrogen (as in the Fiducial run) up to the onset of fragmentation. Following fragmentation, the reduced gas densities in the vicinity of some of the newly formed sinks allow the LW radiation to break out, though it remains shielded near the original sink. 

Because fragmentation occurs only in the LW-Only run, the absence of EUV feedback appears to contribute to destabilizing the disk. The mechanism by which this may occur is as follows: EUV radiation supplies a radiation force that slows down the infalling gas and suppresses accretion shocks relative to the LW-Only run where this radiation is absent. Stronger supersonic shocks therefore form near the protostar in the LW-Only run that induce local heating and suppresses stellar growth. As a result, the disk-to-stellar mass ratio grows leading to a Toomre-unstable disk which fragments. In the following we will discuss each of the above steps. 

EUV radiation, which is implemented in all but the LW-Only run, not only ionizes the gas, but also provides a radiation force that affects its dynamics. Following \citet{Sharda_24}, the radial radiation force per unit volume  can be written as
\begin{equation}
\mathcal{F}_{\mathrm{rad}} = \frac{1}{c} \, \rho \, \kappa \, k_{\mathrm{H}} \, (\mathbf{F} \cdot \hat{\mathbf{r}}) ,
\end{equation}
 where $\rho$ is the gas density, $\kappa$ is the opacity of hydrogen atoms in the relevant UV energy band considered (i.e.\ $13.6 - 15.2 \ \mathrm{eV}$ and $15.2 \ \mathrm{eV} - \infty$), $k_{\mathrm{H}}$ is the number fraction of H, $\mathbf{F}$ is the radiation flux, and $\hat{\mathbf{r}}$ is the radial unit vector in the stellar frame. Within $R\lesssim 200$ AU, where most of the ionizing photon absorption occurs (see Appendix~\ref{app:photon_consumption}), the resulting radiation force per unit volume is $\simeq 10^{-8} \mathrm{~dyn~cm^{-3}}$ far exceeding the gravitational force per unit volume of $\simeq 10^{-14} \mathrm{~dyn~cm^{-3}}$. As described in Appendix A of \citet{Sharda_24} this excess radiation force decelerates inflowing gas and suppresses the formation of strong accretion shocks near the protostar. Ionizing radiation force has also been shown to slow accreting gas inside the H\,\textsc{ii} region \citep[assuming spherical symmetry;][]{Omukai_02}. In the absence of EUV radiation, the LW-Only run lacks this decelerative feedback. Since the LW flux is shielded at this time by atomic hydrogen, it does not effectively contribute to the radiation pressure. 

\begin{figure}[htbp]
    \centering
    \includegraphics[width=\textwidth]{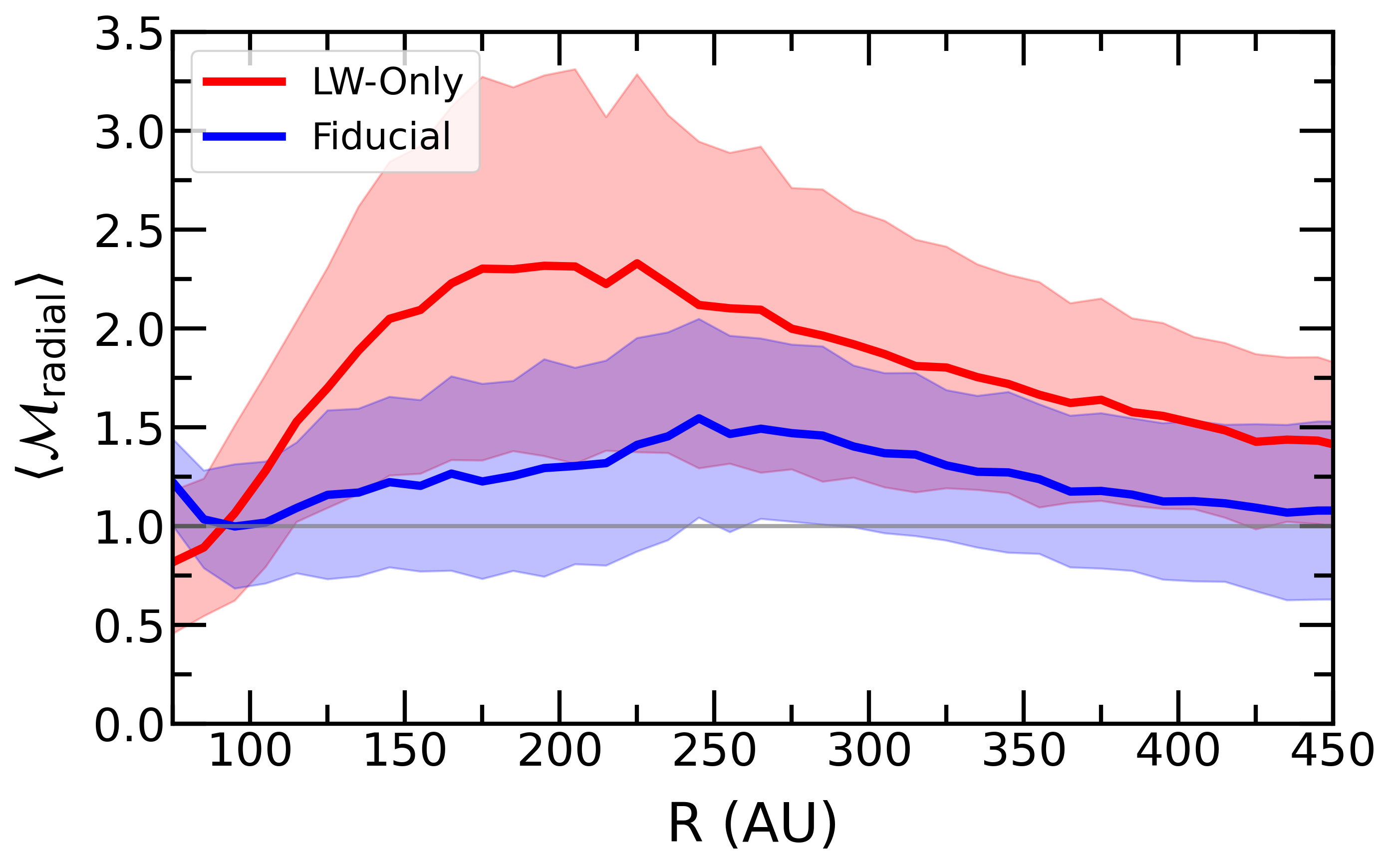}
    \caption{Time-averaged radial Mach number profile of infalling gas for the Fiducial and LW-Only runs computed in spherical shells centered on the protostar and averaged over all simulation outputs up to t = 8 kyr. Solid lines show the mean mass-weighted radial Mach number at each radius, while the shaded regions indicate the temporal standard deviation across snapshots. The LW-Only run exhibits persistently supersonic inflow over a broad radial range, peaking near $\sim$ 200 AU, whereas the Fiducial run remains predominantly transonic or mildly supersonic at all radii.}    
    \label{fig:mach_radial}
\end{figure}

\begin{figure*}[t]
    \centering
    \includegraphics[width=\textwidth]{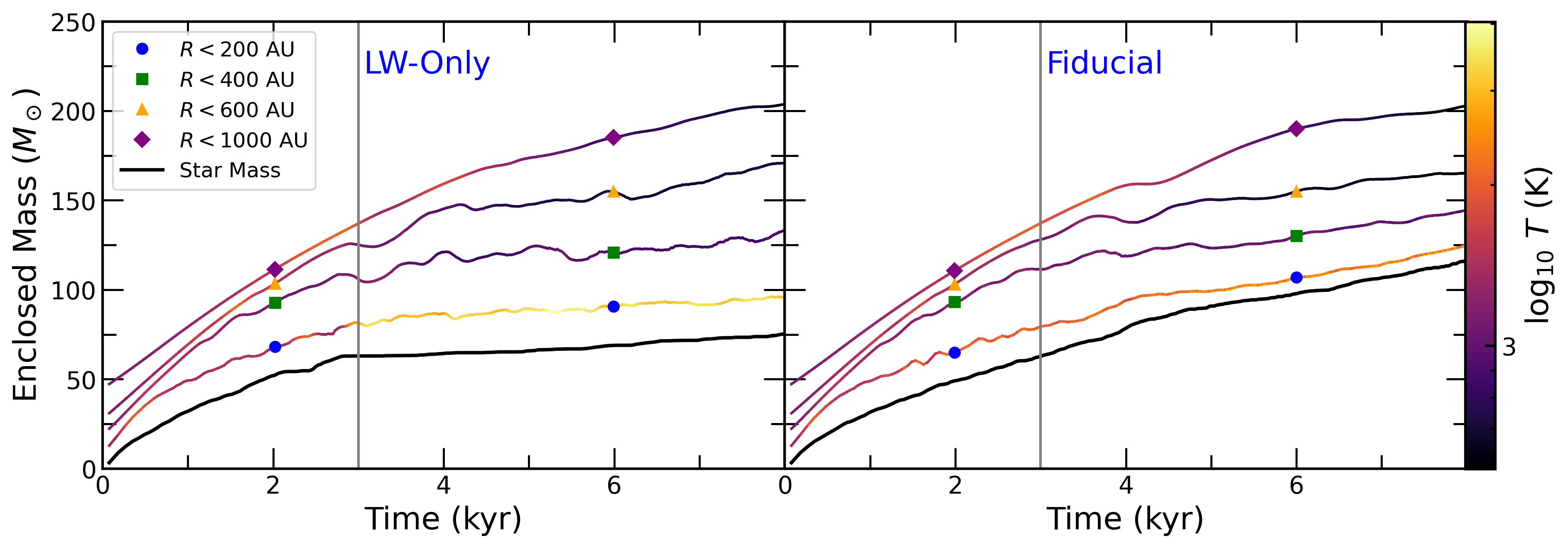}
    \caption{Evolution of the enclosed mass and gas temperature surrounding the protostar during the first 8 kyr after formation, prior to fragmentation in the LW-Only run. The left and right panels show the LW-Only and Fiducial runs, respectively. Each curve shows the total enclosed mass (stellar mass plus gas mass) within spherical radii of 200, 400, 600, and 1000 AU from the protostar, while the curve color encodes the mass-weighted gas temperature in the corresponding spherical shell (0–200 AU: blue circles; 200–400 AU: green squares; 400–600 AU: yellow triangles; 600–1000 AU: purple diamonds). Markers at $t = 2$ kyr and $t = 6$ kyr are included to aid comparison. In the LW-Only run, the innermost region ($R < 200$ AU) undergoes a sharp temperature increase at $t \simeq 3$ kyr, coinciding with a flattening of the stellar mass growth. No comparable temperature jump is observed in the Fiducial run, where accretion proceeds smoothly. Before the temperature increase (e.g. at $t = 2$ kyr), the enclosed-mass profiles are similar in both runs. Afterward (e.g. at $t = 6$ kyr), the outer regions contain comparable mass, while excess mass accumulates preferentially in the hot innermost region of the LW-Only run.}
    \label{fig:enclosed_mass}
\end{figure*}

As a result, inflowing gas in the LW-Only run remains largely unimpeded, reaching supersonic velocities and forming strong accretion shocks near the protostar. In Figure~\ref{fig:mach_radial}, we show the time-averaged radial Mach number, $\langle \mathcal{M}_{\rm radial} \rangle = \langle v_r / c_s \rangle$, of infalling gas computed in spherical shells centered on the protostar and averaged over the same set of simulation outputs used in Figure~\ref{fig:enclosed_mass}. The shaded regions indicate the temporal standard deviation at each radius. We find that at radii of R $\sim 200$ AU the LW-Only run exhibits persistently supersonic inflow, whereas the corresponding flow in the Fiducial run remains transonic to only mildly supersonic.\footnote{With 64 cells per Jeans length, gas crosses shocks faster than it can cool, allowing shock heating to exceed the $\sim 2000$ K threshold for rapid H$_2$ collisional dissociation rather than being artificially radiatively cooled (see \citealt[Appendix A]{Sharda_21}).}

These shocks elevate the gas temperatures in the immediate vicinity of the protostar in the LW-Only run suppressing stellar mass growth. In Figure~\ref{fig:enclosed_mass}, we show the evolution of the mass-weighted gas temperature enclosed within spherical regions during the first 8 kyr following protostar formation, prior to fragmentation in the LW-Only run. Each curve represents the total enclosed mass (including the protostar) as a function of spherical radius R from the sink, color-coded by the mass-weighted temperature in each spherical shell. In the LW-Only run (left panel), the temperature rises sharply to temperatures $>$ 2,000 K, the dissociation threshold of H$_2$, at t $\simeq$ 3 kyr within the inner region (i.e., R $<$ 200 AU; blue circle), precisely the same region where we saw elevated shocks in Figure~\ref{fig:mach_radial}. This temperature increase coincides with the flattening of the stellar mass curve (black line). In contrast, no comparable temperature increase is observed in the Fiducial run (right panel), and the protostar continues to accrete smoothly throughout this time. This suggests that the localized temperature increase suppresses the mass growth of the protostar by thermally stabilizing the surrounding gas and impeding accretion from the immediate environment, as discussed in Section \ref{subsec:The Impact of H Shielding}.

The temperature increase in the LW-Only run is primarily a local effect and does not significantly modify mass transport from large to intermediate scales. This is evident from a comparison of the enclosed-mass profiles in the LW-Only and Fiducial runs shown in Figure~\ref{fig:enclosed_mass}, both before and after the temperature enhancement at $t \simeq 3$~kyr. For example, at $t = 2$~kyr, the relative spacing between adjacent enclosed-mass curves in the two runs -- aside from the innermost curve -- closely matches that at $t = 6$~kyr, following the temperature increase. This similarity indicates that the temperature enhancement primarily affects the gas in the immediate vicinity of the protostar, while mass inflow from larger radii remains largely unchanged. The primary difference between the two runs is seen within the innermost region ($R \lesssim 200$ AU; blue circle). In the LW-Only run, as opposed to the Fiducial run, a large reservoir of gas accumulates near the protostar, as indicated by the increased separation between the innermost enclosed-mass curve and the stellar-mass curve. This gas is significantly hotter and therefore provides enhanced thermal pressure support, reducing its ability to accrete efficiently, thereby decreasing the protostellar mass in the LW-Only run compared to the Fiducial run. 

The suppression of protostellar growth in the LW-Only run leads to an increasingly high disk-to-stellar mass ratio. As already suggested by Figure~\ref{fig:enclosed_mass}, the LW-Only and Fiducial runs contain comparable amounts of gas at all but the innermost radii ($R \lesssim 200$ AU; blue circle), while the LW-Only run hosts a substantially less massive protostar. As a result, the disk-to-stellar mass ratio in the LW-Only run grows rapidly with time. By $t \simeq 11$ kyr, immediately prior to fragmentation, this ratio reaches $\sim$ 2:1, where we define the disk as gas with number density $n > 10^{9}\ \mathrm{cm^{-3}}$, a threshold that visually separates the dense, rotationally supported midplane structure from the lower-density infalling envelope. Such elevated disk-to-stellar mass ratios are known to promote gravitational instability and fragmentation when they exceed unity \citep[e.g.,][]{Kratter_10, Kimura_21}. In contrast, the Fiducial run at this time, and at earlier times when its stellar mass is comparable to that of the LW-Only run at $t \simeq 11$~kyr, maintains a disk-to-stellar mass ratio below unity and does not fragment. Thus, the effect of the absence of EUV radiation on fragmentation is indirect. It suppresses protostellar mass by allowing strong shocks to develop near the sink that impede accretion. The smaller protostellar mass increases the disk-to-stellar mass ratio and in turn weakens the gravitational stability of the disk at larger radii (the second sink forms $\sim 500$ AU from the primary sink).

\begin{figure}[htbp]
    \centering
    \includegraphics[width=\textwidth]{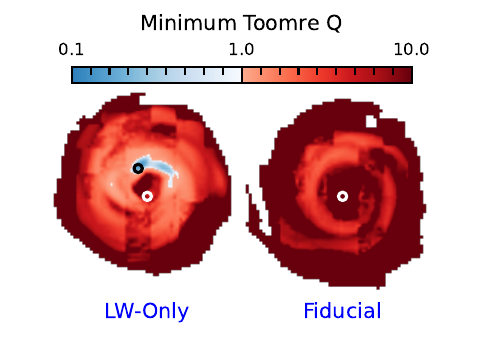}
    \caption{Projection map of the minimum Toomre \(Q\) parameter in the disk for the LW-Only run (left) at the snapshot immediately preceding fragmentation and for the Fiducial run (right) at the snapshot closest to this same time. White circles mark the locations of existing sink particles, while the black circle in the LW-Only panel indicates the location of the sink particle that forms in the subsequent snapshot, $\sim 500$ AU from the existing sink.}
    \label{fig:toomre}
\end{figure}

We can further assess the gravitational stability of the disk by evaluating the Toomre parameter $Q$ which provides a criterion for disk stability. While the Toomre criterion is formally derived for infinitesimally thin isothermal disks, it remains valid to within a factor of order unity for geometrically thick disks as in our simulation \citep[e.g.,][]{Wang_10}. When $Q \gtrsim 1 $ shear by differential rotation and gas pressure prevent collapse and the disk is stable, but when $Q \lesssim 1 $ gravitational forces dominate, leading to disk instability and favoring the development of spiral arms that transport mass inward and angular momentum outward. The Toomre parameter is given by
\begin{equation}
Q = \frac{c_{\rm{s}} \kappa}{\pi G \Sigma},
\end{equation}
where $\Sigma$ is the gas surface density. We compute $\Sigma$ and $c_{\rm{s}}$ locally for each gas cell and use the orbital frequency $\Omega = \sqrt{G M(<r)/r^3}$ as a substitute for the epicyclic frequency $\kappa$, an appropriate substitution for a Keplerian disk. We verify that our disk is approximately Keplerian beyond $r \gtrsim 100$ AU. 

In Figure~\ref{fig:toomre}, we show two-dimensional projections of the minimum Toomre \(Q\) in the disk, centered on the primary sink particle (white circle). The left panel shows the LW-Only run in the snapshot immediately preceding fragmentation (\(t \simeq 11\,\mathrm{kyr}\)), while the right panel shows the Fiducial run at the snapshot closest to this same time. In the LW-Only run, the disk is largely marginally stable, with \(Q \sim 1\) over most radii. An extended annular region at a radius of \( r\sim 500\,\mathrm{AU}\) exhibits \(Q < 1\), indicating local gravitational instability. A second sink particle forms in the subsequent snapshot within this unstable annulus and is marked by the black circle in the figure. In contrast, the Fiducial run, which does not undergo fragmentation, shows no extended regions with \(Q \lesssim 1\), remaining gravitationally stable throughout the disk.

\section{Discussion \& Conclusion}
\label{sec:discussion}

\subsection{Confinement of Ionized Region}
\label{subsec:confinement}

The compact H\,\textsc{ii} region seen in our simulations likely reflects a stage of dynamical evolution in Pop III star-forming disks. The dynamical expansion of the H\,\textsc{ii} region depends not only on the ionization output and recombination rate of the dense inner-region gas, but on the gravity of the star (which binds the ionized gas), the density of the accretion flow (which increases recombinations), and the infalling ram pressure (which opposes thermal expansion) \citep[e.g.,][]{Keto_03,McKee_08, Galv_11, Stacy_16, Lund_19}. These complex and nonlinear effects may conspire to keep primordial H\,\textsc{ii} regions compact, analogous to the variable ultracompact and hypercompact H\,\textsc{ii} regions observed around present-day massive stars (see \citealt{Martini_26}, and references therein). This can facilitate extended periods of accretion onto the star and produce very/extremely massive stars. Three-dimensional simulations that resolve both the inner structure and the long-term evolution of H\,\textsc{ii} regions are therefore needed to determine whether, and to what extent, this can affect the final mass of Pop III stars.

Magnetic fields may also impact the evolution of the H\,\textsc{ii} region. Magnetic fields alter the gas distribution around the protostar, reducing the central density while producing denser, more extended structures at larger radii \citep[e.g.,][]{Sharda_24, Sadanari_24}. Such redistribution may have opposing effects on H\,\textsc{ii} region evolution: lower-density channels near the star can facilitate escape channels for ionizing photons, accelerating expansion, while dense material farther out can increase recombinations and ram pressure, delaying or confining expansion. In this sense, magnetic fields may introduce anisotropic escape pathways analogous to those produced by rotation, where departures from spherical symmetry allow expansion in some directions while maintaining shielding in others \citep{Omukai_02, McKee_08}. Magnetic fields may also regulate H\,\textsc{ii} region morphology by suppressing fragmentation, slowing expansion perpendicular to the field lines \citep[e.g.,][]{Krumholz_07}, and constraining expansion when the ionized-gas thermal pressure becomes comparable to the magnetic pressure of the surrounding gas \citep{Peters_11}. 

Turbulence may have a similar two-sided effect on the H\,\textsc{ii} region. By modifying the density field, it can both open low-density escape channels and create dense structures that enhance recombinations. It may also generate ram pressure against the expanding ionized gas \citep[e.g.,][]{Geen_15}. Similar behavior is suggested by \citet{Sugimura_23}, who found that Pop III clouds with higher accretion rates, and correspondingly larger turbulent velocities, develop more compact photoionized regions (see their Figure 3). In our simulations, the vertically extended, striated turbulent density field in Figure~\ref{fig:fields_slice} may likewise contribute to the small fluctuations in the H\,\textsc{ii} region seen in Figure~\ref{fig:fields_slice} and quantified in Figure~\ref{fig:HII}. In future work, we will quantify the relative importance of these confinement mechanisms in regulating the expansion and morphology of primordial H\,\textsc{ii} regions.

Because H\,\textsc{ii} region breakout can terminate accretion onto the protostar \citep[e.g.,][]{Hosokawa_16, Sugimura_23, Toyouchi_23}, understanding how magnetic fields, turbulence, and accretion jointly affect the density structure and ionizing output is essential for robustly determining the final masses of Pop III stars.

\subsection{\texorpdfstring{Ly$\alpha$}{Ly-alpha} Radiation Pressure}

While our simulations include the effects of ionizing and LW dissociation feedback, we do not include the momentum imparted by resonantly scattered Ly$\alpha$ photons due to the complications involved in their modeling. This could be especially pertinent in the dust-free (where dust destruction of Ly$\alpha$ photons is absent) and high $N_{\rm HI}$ columns realized in the vicinity of the H\,\textsc{ii} regions, which likely lead to radiative forces well in excess of other radiative feedback channels \citep[e.g.,][]{Nebrin_25,Menon_26}. This additional force may facilitate the breakout of the H\,\textsc{ii} region \citep[e.g.,][]{McKee_08,Jaura_22}. On the other hand, high $N_{\rm HI}$ also imply deep gravitational potential wells, and the outcome therefore depends on the competition between Ly$\alpha$ forces and gravity. In addition, at these high densities/columns Ly$\alpha$ emission can be suppressed through collisions inducing 2p$\rightarrow$2s transitions and subsequent 2-photon emission, acting to saturate the momentum imparted \citep[e.g.,][]{Nebrin_25}. 

To provide some intuition on whether Ly$\alpha$ pressure could change the evolution of the H\,\textsc{ii} region in our simulations, we quantify the Ly$\alpha$-gravity competition by computing the critical H\,\textsc{i} column below which the Ly$\alpha$ force can overcome the gravitational force. For a neutral shell of surface density $\Sigma_{\rm HI}$, with
$M_{\rm shell}=4\pi r^2\Sigma_{\rm {HI}}$, the Ly$\alpha$ acceleration is

\begin{equation} 
a_\alpha \simeq \frac{M_F L_\alpha}{c\, M_{\rm shell}}, 
\end{equation} 
 where $M_F$ is the force multiplier caused by resonant scattering and $L_\alpha$ is the Ly$\alpha$ luminosity. Comparing this with the gravitational acceleration $a_{\rm grav}=GM_*/r^2$, we can define a critical surface density below which Ly$\alpha$ pressure can overcome the gravitational attraction of the star: 

\begin{equation} 
\Sigma_{\rm{HI}, \rm{max}} = \frac{M_F L_\alpha}{4\pi c G M_* }. 
\end{equation} 

Converting this to H\,\textsc{i} column by using $\Sigma_{\rm HI}=m_{\rm H}N_{\rm HI}$ and estimating $L_\alpha$ by assuming that a fraction $f_\alpha=2/3$ of absorbed ionizing photons are converted into Ly$\alpha$ photons, so that $L_\alpha\simeq f_\alpha h\nu_\alpha Q_{\rm EUV}$, with $h\nu_\alpha=10.2~{\rm eV}$, we get

\begin{equation}
N_{\rm HI,\max} \simeq 7\times10^{24} \left(\frac{M_F}{50}\right) \left[ \frac{Q_{\rm EUV}/M_*} {10^{48}{\rm s^{-1}}M_\odot^{-1}} \right] {\rm cm}^{-2}.
\end{equation}
 where we set the fiducial $M_F$ to 50 as suggested by \citet{nebrin_26}. Taking representative values of $Q_{\rm EUV}/M_*\sim 10^{48}\ {\rm s^{-1}}\ M_\odot^{-1}$, we find that typical polar H\,\textsc{i} columns of $N_{\rm HI} \simeq 10^{26-27} \mathrm{cm}^{-2}$ are larger than the critical column estimated above, implying that Ly$\alpha$ radiation pressure is sub-Eddington with respect to the stellar gravitational force. Since this comparison neglects the additional ram, turbulent, and magnetic confinement, it should be regarded as an upper limit on the ability of Ly$\alpha$ pressure to drive breakout. We therefore do not expect Ly$\alpha$ pressure to qualitatively change the conclusion that the H\,\textsc{ii} region remains confined under the conditions considered here. That being said, nonlinear coupling between the various feedback pathways may lead to deviations from the simplified analysis above, and on-the-fly treatments of Ly$\alpha$ radiation pressure \citep[e.g.,][]{nebrin_26} would be required to confirm these expectations.

\subsection{Resolution}
\label{subsec:resolution}

A caveat to this work is that the H\,\textsc{i} shielding in the simulations is controlled by unresolved structure. In the non-fragmenting runs, the innermost cells lying within the sink-accretion radius provide the high H\,\textsc{i} column that shields the LW radiation. Removing these innermost cells would typically lower the H\,\textsc{i} column by several orders of magnitude. We nevertheless expect H\,\textsc{i} shielding to remain important at higher resolution for two reasons. First, we expect high columns at higher resolution. For instance, the simulations of \citet{Jaura_22} have a resolution of 1 AU and column densities of $N_{\rm HI} \sim 10^{25}\,\mathrm{cm}^{-2}$. Second, because the innermost gas is expected to be subsonic, it is unlikely to develop low-column channels through which LW radiation can escape. Thus, while unresolved structure provides the H\,\textsc{i} shielding in this work, we expect its impact to remain important at higher resolution.

The fragmentation of the disk and mass of the stars is also resolution dependent. Higher resolution promotes fragmentation by creating finer and sharper density structures. This typically results in a larger number of lower-mass sinks, while leaving the total mass accreted into sinks largely unchanged \citep[e.g.,][]{Prole_22,Park_23,Sugimura_23}. In contrast, in simulations where the cloud is restricted to forming a single sink, higher resolution can lead to more efficient transport of mass and angular momentum, thereby increasing the final stellar mass \citep{Hosokawa_16}.\footnote{We note that resolution is not solely determined by the maximum refinement level; the number of resolution elements maintained at that level is equally critical. A useful metric is the number of cells per Jeans length, which measures how well gravitational collapse is resolved.} The disk-to-stellar mass ratio has also been shown to depend on resolution \citep[e.g.,][]{Kimura_21}. Consistent with these trends, increasing the maximum spatial resolution in our Fiducial run from $\Delta x = 30\,\mathrm{AU}$ to $\Delta x = 7.5\,\mathrm{AU}$ leads to disk fragmentation \citep{Sharda_25}. Because fragmentation alters the disk geometry and density structure near the protostar, such resolution dependent changes may also influence the expansion of the H\,\textsc{ii} region.

The morphology and confinement of the H\,\textsc{ii} region may be further influenced by numerical resolution. First, the Str\"omgren radius in our simulations is only marginally resolved, which may affect the strength and spatial coupling of EUV feedback \citep[e.g.,][]{Susa_13,Stacy_16}. Higher-resolution simulations will be required to test convergence and determine whether confinement persists under improved resolution. Second, because the vertical scale height of the inner disk is not resolved, the peak gas density in the immediate vicinity of the protostar may not be accurately captured, leading to uncertainties in the column density encountered by ionizing photons along polar directions and in the local recombination time, $t_{\rm rec} = (\alpha_{\rm B} n_e)^{-1}$. At the same time, numerical averaging may artificially smooth the vertical density structure of the disk. Such effects may alter both the replenishment of dense gas near the sink and the efficacy of low-density escape channels, influencing the confinement of the H\,\textsc{ii} region.

The confinement of the H\,\textsc{ii} region may also depend on the method used to inject radiation. \citet{Jaura_22} showed that when ionizing radiation is injected inside the sink rather than at its surface, the radiation can be trapped by dense gas within the sink for at least the first $\sim 20$ kyr. Most simulations that form expanded H\,\textsc{ii} regions inject radiation at the sink surface \citep[e.g.,][]{Hosokawa_16,Park_23}, although some models explicitly account for radiative transfer within the sink itself \citep[e.g.,][]{Sugimura_20}. In our simulations, ionizing radiation is injected below the sink surface, which may lead to enhanced local absorption near the protostar and thereby contribute to the delayed expansion and confinement of the H\,\textsc{ii} region. However, since the density structure surrounding Pop III protostars is unknown, the correct injection model is still uncertain \citep[e.g.,][]{Sugimura_23}.

\subsection{Comparison with Other Studies}

The relative roles of EUV and FUV radiation in regulating Pop III protostellar growth remain uncertain. A number of numerical studies have found that LW-driven heating can substantially modify the thermal and density structure of gas near the protostar and ultimately halt protostellar growth when the H\,\textsc{ii} region remains confined near the sink \citep[e.g.,][]{Umemura_12, Stacy_12, Susa_13, Stacy_16}. Consistent with these numerical results, we find that LW radiation drives a sharp decline in accretion at $\sim$ 25 kyr and fully quenches accretion by $\sim$ 55 kyr in our No-Hshield run. More recently, \citet{Park_23} also reported that FUV-only feedback can halt accretion in a manner comparable to EUV-only feedback.

Other studies, on the other hand, have concluded that FUV feedback plays a secondary role compared to EUV feedback. Using simulations on a spherical grid, \citet[][see their Figure 8]{Hosokawa_16} found that LW feedback produces only modest changes to the accretion rate, whereas EUV radiation is required to halt accretion. Similarly, \citet[][see their Figure 13]{Sugimura_23} and \citet[][in the context of low-metallicity, $\mathrm{[Z/H]}=-4$, see their Figure 5]{Chon_24}, show that LW radiation can suppress stellar growth (although to a greater degree than \citealt{Hosokawa_16}), but both conclude that EUV radiation is required to halt accretion. This contrasts with our results, where we find that LW radiation, when not shielded by H\,\textsc{i}, can halt accretion. Differences in the treatment of H\,\textsc{i} shielding may contribute to the discrepancies among studies.

Our findings regarding the impact of H\,\textsc{i} shielding on stellar growth resemble those of \citet{Park_23}. In their simulations, H\,\textsc{i} absorption in the LW band is incorporated through a fixed effective cross-section within an optically thin framework, together with a cell-based column approximation. They show that atomic hydrogen can provide opacity comparable to H$_2$ self-shielding, substantially weakening LW feedback and increasing the total mass accreted onto sinks (see their Appendix~A4). In contrast, we determine the H\,\textsc{i} attenuation by explicitly integrating the hydrogen column density along rays from each sink through the adaptive mesh, allowing the LW suppression to depend directly on the resolved $N_{\rm HI}$. We find that this dynamically computed attenuation weakens LW feedback and enhances stellar mass growth to a greater degree than in \citet{Park_23}, highlighting that the quantitative impact of H\,\textsc{i} shielding can depend on how it is modeled.

\subsection{Role of Environment} 
In the canonical picture, Pop III star formation occurs in minihalos of mass $\sim 10^{5–6}\ \mathrm{M_\odot}$, where H$_2$ cooling enables gas to contract to high densities despite the absence of metals \citep[e.g.,][]{Haiman_96,Tegmark1997}. Cosmological baryon–dark matter streaming velocities, coherent over megaparsec scales, provide one of the most important modifications to this environment \citep{Tes+10a}. These supersonic flows increase turbulence \citep{Chen_25} and suppress gas accretion into the smallest halos \citep[e.g.,][]{Fialkov+11}, thereby delaying Pop III formation in some regions \citep[e.g.,][]{Schauer+19, hirano_25}. They also allow dense gas structures to form outside the virial radius of any minihalo \citep[e.g., so-called Supersonically-Induced Gas Objects or SIGOs,][]{Naoz+14}. Recent simulations have shown that in high-streaming patches, gas can condense into SIGOs, which are dense, metal-free clumps largely devoid of dark matter \citep[e.g.,][]{Naoz+12,Chiou+18, Popa+15,Williams_2024, Williams+23}.  These structures represent an alternative pathway to Pop III star formation, potentially forming compact stellar clusters in environments where typical minihalo collapse is inhibited. 

While minihalos provide the canonical environment for Pop III star formation, the physical conditions within SIGOs differ in several important ways and may lead to distinct star-formation pathways. In minihalos, gravitational collapse is regulated by the depth of the dark matter potential well, with gas able to cool via H$_2$ once densities are high enough for effective self-shielding, typically producing a small number of massive protostars \citep[e.g.,][]{Yoshida_08, Hirano+14}. By contrast, the absence of a dominant dark matter potential means that SIGOs rely entirely on gas self-gravity and cooling to reach the densities required for fragmentation, often resulting in more compact, pressure-supported structures \citep{Lake+23b,Lake+24b}. Simulations suggest that SIGOs may achieve higher central densities and more efficient H$_2$ cooling than comparable gas in low-mass halos suppressed by streaming, potentially generating denser, more cluster-like Pop III star-forming regions \citep[e.g.,][]{Chiou+21, Williams_25}. Additionally, because SIGOs form outside halo centers, their exposure to external LW radiation backgrounds differs from that of minihalo cores, and their ability to self-shield may vary accordingly.

\subsection{Conclusion}

We have analyzed a suite of radiation-magnetohydrodynamics simulations from the \textsc{Popsicle} project to quantify how H\,\textsc{i} shielding of LW radiation and the inclusion of different radiative feedback affect the growth of Pop III stars. Our primary results can be summarized as follows:

\begin{enumerate}
    
\item In the absence of H\,\textsc{i} shielding, LW feedback significantly suppresses accretion and limits stellar growth. LW heating near the protostar reduces H$_2$ abundance, raises gas temperatures, and ultimately shuts off accretion by $\sim$ 55 kyr, even while the H\,\textsc{ii} region remains confined.

\item Including H\,\textsc{i} shielding introduces an additional opacity to LW photons in the immediate protostellar environment, particularly in the polar direction where the build-up of large H\,\textsc{i} columns shields the transmitted LW flux where H$_2$ self-shielding alone is insufficient. As a result, neglecting H\,\textsc{i} shielding lowers the stellar mass at the final output by 22\% relative to the Fiducial run, demonstrating that this omission overestimates the effectiveness of LW feedback in regulating Pop~III stellar growth.

\item In runs with EUV feedback, the H\,\textsc{ii} region remains confined to $\sim 100$~AU measured outward from the sink accretion radius despite high ionizing luminosities. Dense, gravitationally bound gas sustains high recombination rates and prevents sustained pressure-driven expansion. Turbulence and magnetic fields may also contribute to its confinement.

\end{enumerate}

Taken together, these results show that H\,\textsc{i} shielding of LW radiation can qualitatively alter the final masses of Pop III stars by weakening radiative feedback in the immediate vicinity of the protostar. Because the final stellar mass determines the fate of Pop III stars -- ranging from core-collapse supernovae to direct collapse black holes -- including atomic hydrogen shielding will help better predict their fates. In our simulations this effect is sufficiently strong to shift the stellar remnant channel: the sink mass in the No-Hshield run reaches $218\ \mathrm{M_\odot}$, consistent with a pair-instability supernova progenitor \citep[][]{Heger_10}, whereas in the Fiducial run it reaches $279\ \mathrm{M_\odot}$, likely collapsing directly into a black hole. Although our simulations are not definitive, this comparison shows that H\,\textsc{i} shielding may influence not only stellar masses, but the fate of Pop III stars.

More broadly, LW radiation has also been proposed as a key ingredient in scenarios for primordial supermassive black hole formation. This has been extensively explored across a range of environments, including within the host itself \citep{Dunn_18, Chiaki_23, Sullivan_25} or in neighboring halos exposed to a local external irradiation \citep[e.g.,][]{Dijkstra_14, Visbal_14b, Regan_16, Chon_25,vanVeenen_2025}. Our results indicate that H\,\textsc{i} shielding can weaken LW feedback on small scales, potentially modifying the conditions under which such black hole seeding can occur. 


\vspace{1em}
{\bf Acknowledgments.} 
We thank Zoltan Haiman for useful discussions on H\,\textsc{i} shielding. A.C., S.M., B.B., and S.N. acknowledge support from NASA grant 80NSSC24K0773  (ATP-23 -- ATP23-0149). B.B. acknowledges NSF grant AST-2407877.
B.B. is grateful for the generous support by the David and Lucile Packard Foundation and the Alfred P. Sloan Foundation. B.B. thanks the Center for Computational Astrophysics (CCA) of the Flatiron Institute and the Mathematics and Physical Sciences (MPS) division of the Simons Foundation for support. The Flatiron Institute is supported by the Simons Foundation. P.S. is supported by the Leiden University Oort Fellowship and the International Astronomical Union -- Gruber Foundation Fellowship.
\software{ {\tt FLASH} \citep{Fryxell_00}, {\tt{VETTAM}} \citep{Menon_22}, matplotlib \citep{Matplotlib}, numpy \citep{numpy}, scipy \citep{SciPy}, and {\tt yt} \citep{Turk+11}.}

\appendix

\section{EUV Photon Consumption}
\label{app:photon_consumption}

\begin{figure}[htbp]
    \centering
    \includegraphics[width=\columnwidth]{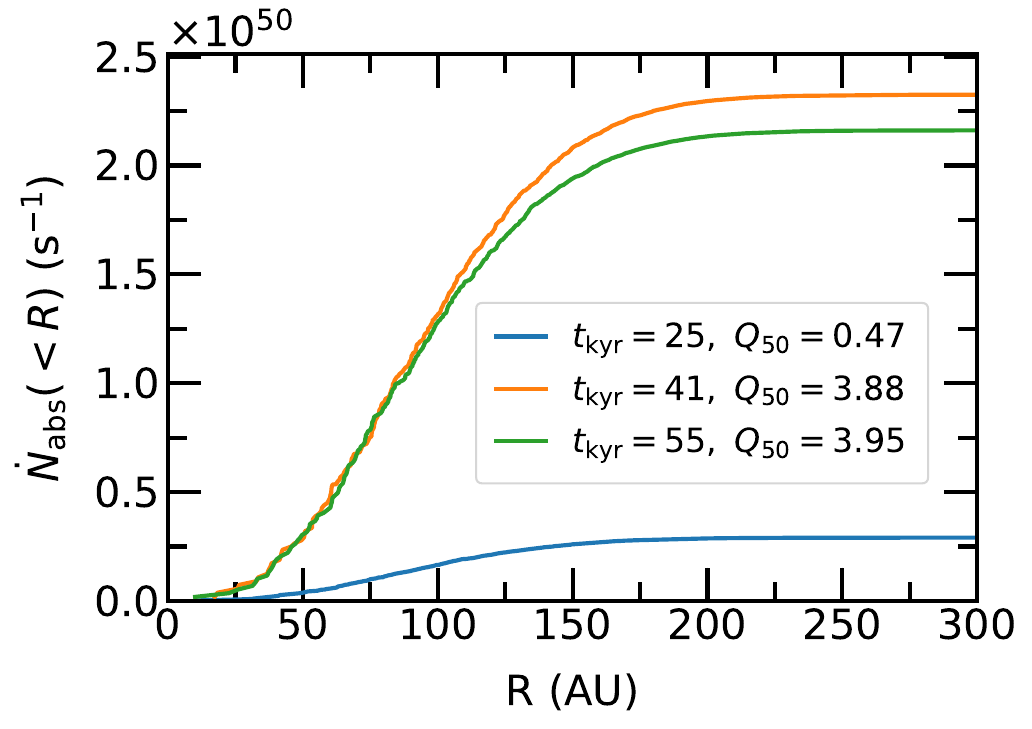}
    \caption{Cumulative EUV photon absorption rate,
    $\dot{N}_{\rm abs}(<R)$ around the sink. The plotted quantity includes photon consumption by both H\,\textsc{i} and H$_2$. The legend gives the sink age and $Q_{50}\equiv Q_{\rm EUV}/10^{50}\ {\rm s}^{-1}$ for each snapshot.}
    \label{fig:EUV_photon_absorption}
\end{figure}

We quantify where ionizing photons are absorbed by computing the cumulative EUV photon absorption as

\begin{equation}
    \dot{N}_{\rm abs}(<R)
    =
    \dot{N}_{\rm abs,H\,\textsc{i}}(<R)
    +
    \dot{N}_{\rm abs,H_2}(<R).
\end{equation}
 Figure~\ref{fig:EUV_photon_absorption} shows this quantity for the same three snapshots in Figure~\ref{fig:fields_slice}. For all three times, we find that the EUV consumption rate rises over the inner region and saturates at $\sim 200$ AU. The H\,\textsc{ii} region is therefore regulated by the unresolved inner disk and the smallest resolved scales.

\clearpage
\bibliography{cosmo}

\end{document}